\documentclass[onecolumn,12pt]{IEEEtran}
\usepackage{algorithmic}
\usepackage{amsmath}
\usepackage{amssymb}
\usepackage{amsfonts}
\usepackage{array}
\usepackage{epsfig}
\usepackage{setspace}
\usepackage{cite}
\usepackage{calc}
\usepackage{color}
\usepackage{subfig}
\usepackage{paralist}

\title{Interference Alignment in Regenerating Codes for Distributed Storage: Necessity and Code Constructions}

\author{Nihar~B.~Shah, K.~V.~Rashmi, P.~Vijay~Kumar, {\em Fellow, IEEE}, and Kannan~Ramchandran, {\em Fellow, IEEE}
\thanks{The material in this paper was presented in part at the ITA Workshop at UCSD, 2010, in part at the IEEE Information Theory Workshop, Cairo, Egypt, January 2010 and in part at Allerton 2009.}
\thanks{Nihar~B.~Shah, K.~V.~Rashmi and P.~Vijay~Kumar are with the Department of Electrical
Communication Engineering, Indian Institute of Science, Bangalore,
560 012 India (email: \{nihar,rashmikv,vijay\}@ece.iisc.ernet.in).  P. Vijay Kumar is also an adjunct faculty member of the Electrical Engineering Systems Department at the University of Southern California, Los Angeles, CA 90089-2565. }
\thanks{Kannan Ramchandran is with the Department of Electrical Engineering and Computer Science, University of California at Berkeley, Berkeley, CA 94720 USA (e-mail: kannanr@eecs.berkeley.edu).}
}

\newtheorem{defn}{Definition}
\newtheorem{thm}{Theorem}

\newtheorem{lem}[thm]{Lemma}
\newtheorem{cor}[thm]{Corollary}
\newtheorem{note}{Remark}

\newtheorem{pty}{Property}

\newcommand{\beq}{\begin{equation}}
\newcommand{\eeq}{\end{equation}}
\newcommand{\bea}{\begin{eqnarray}}
\newcommand{\eea}{\end{eqnarray}}
\newcommand{\bean}{\begin{eqnarray*}}
\newcommand{\eean}{\end{eqnarray*}}
\newcommand{\bit}{\begin{itemize}}
\newcommand{\eit}{\end{itemize}}
\newcommand{\ben}{\begin{enumerate}}
\newcommand{\een}{\end{enumerate}}
\newcommand{\blem}{\begin{lem}}
\newcommand{\elem}{\end{lem}}
\newcommand{\bthm}{\begin{thm}}
\newcommand{\ethm}{\end{thm}}
\newcommand{\bpf}{\begin{proof}}
\newcommand{\epf}{\end{proof}}

\begin{document}

\maketitle
\thispagestyle{empty}
\bibliographystyle{ieeetran}

\begin{abstract}
Regenerating codes are a class of recently developed codes for distributed storage that, like Reed-Solomon codes, permit data recovery from any arbitrary $k$ of $n$ nodes.  However regenerating codes possess in addition, the ability to repair a failed node by connecting to any arbitrary $d$ nodes and downloading an amount of data that is typically far less than the size of the data file. This amount of download is termed the repair bandwidth. Minimum storage regenerating~(MSR) codes are a subclass of regenerating codes that require the least amount of network storage; every such code is a maximum distance separable~(MDS) code. Further, when a replacement node stores data identical to that in the failed node, the repair is termed as \textit{exact}.

The four principal results of the paper are  (a) the explicit construction of a class of MDS codes for $d  =  n-1 \geq 2k-1$ termed the MISER code, that achieves the cut-set bound on the repair bandwidth for the exact-repair of systematic nodes, (b) proof of the necessity of interference alignment in exact-repair MSR codes, (c) a proof showing the impossibility of constructing linear, exact-repair MSR codes for $d< 2k-3$ in the absence of symbol extension, and (d) the construction, also explicit, of MSR codes for $d=k+1$.  Interference alignment~(IA) is a theme that runs throughout the paper: the MISER code is built on the principles of IA and IA is also a crucial component to the non-existence proof for $d < 2k-3$. To the best of our knowledge, the constructions presented in this paper are the first, explicit constructions of regenerating codes that achieve the cut-set bound.
\end{abstract}

\section{Introduction}\label{sec:intro}

In a distributed storage system, information pertaining to a data file is dispersed across nodes in a network in such a manner that an end-user~(whom we term as a data-collector, or a DC) can retrieve the data stored by tapping into neighboring nodes.  A popular option that reduces network congestion and that leads to increased resiliency in the face of  node failures, is to employ erasure coding, for example by calling upon maximum-distance-separable~(MDS) codes such as Reed-Solomon~(RS) codes.

Let $B$ be the total number of message symbols, over a finite field $\mathbb{F}_q$ of size $q$.  With RS codes, data is stored across $n$ nodes in the network in such a way that the entire data can be recovered by a data-collector by connecting to any arbitrary $k$ nodes, a process of data recovery that we will refer to as \textit{reconstruction}.  Several distributed storage systems such as RAID-6, OceanStore~\cite{oceanstore} and Total~Recall~\cite{totalRecall} employ such an  erasure-coding option.

Upon failure of an individual node, a self-sustaining data storage network must necessarily possess the ability to repair the failed node. An obvious means to accomplish this, is to permit the replacement node to connect to any  $k$ nodes, download the entire data, and extract the data that was stored in the failed node.  For example, RS codes treat the data stored in each node as a single symbol belonging to the finite field $\mathbb{F}_q$. When this is coupled with the restriction that individual nodes perform linear operations over $\mathbb{F}_q$, it follows that the smallest unit of data that can be downloaded from a node to assist in the repair of a failed node (namely, an $\mathbb{F}_q$ symbol), equals the amount of information stored in the node itself.  As a consequence of the MDS property of an RS code, when carrying out repair of a failed node, the replacement node must necessarily collect data from at least $k$ other nodes.  As a result, it follows that the total amount of data download needed to repair a failed node can be no smaller than $B$, the size of the entire file.   But clearly, downloading the entire $B$ units of data in order to recover the data stored in a single node that stores only a fraction of the entire data file is wasteful, and raises the question as to whether there is a better option. Such an option is provided by the concept of a \emph{regenerating code} introduced by Dimakis et~al.~\cite{DimKan1}.

Regenerating codes overcome the difficulty encountered when working with an RS code by working with codes whose symbol alphabet is a vector over $\mathbb{F}_q$, i.e., an element of $\mathbb{F}_q^{\alpha}$ for some parameter $\alpha > 1$.  Each node stores a vector symbol, or equivalently stores $\alpha$ symbols over $\mathbb{F}_q$.  In this setup, it is clear that while maintaining linearity over $\mathbb{F}_q$, it is possible for an individual node to transfer  a fraction of the data stored within the node.

\begin{figure}[t]
\centering
\subfloat[]{\includegraphics[trim=0in 1.81in 7in 0in, clip, width=0.3\textwidth]{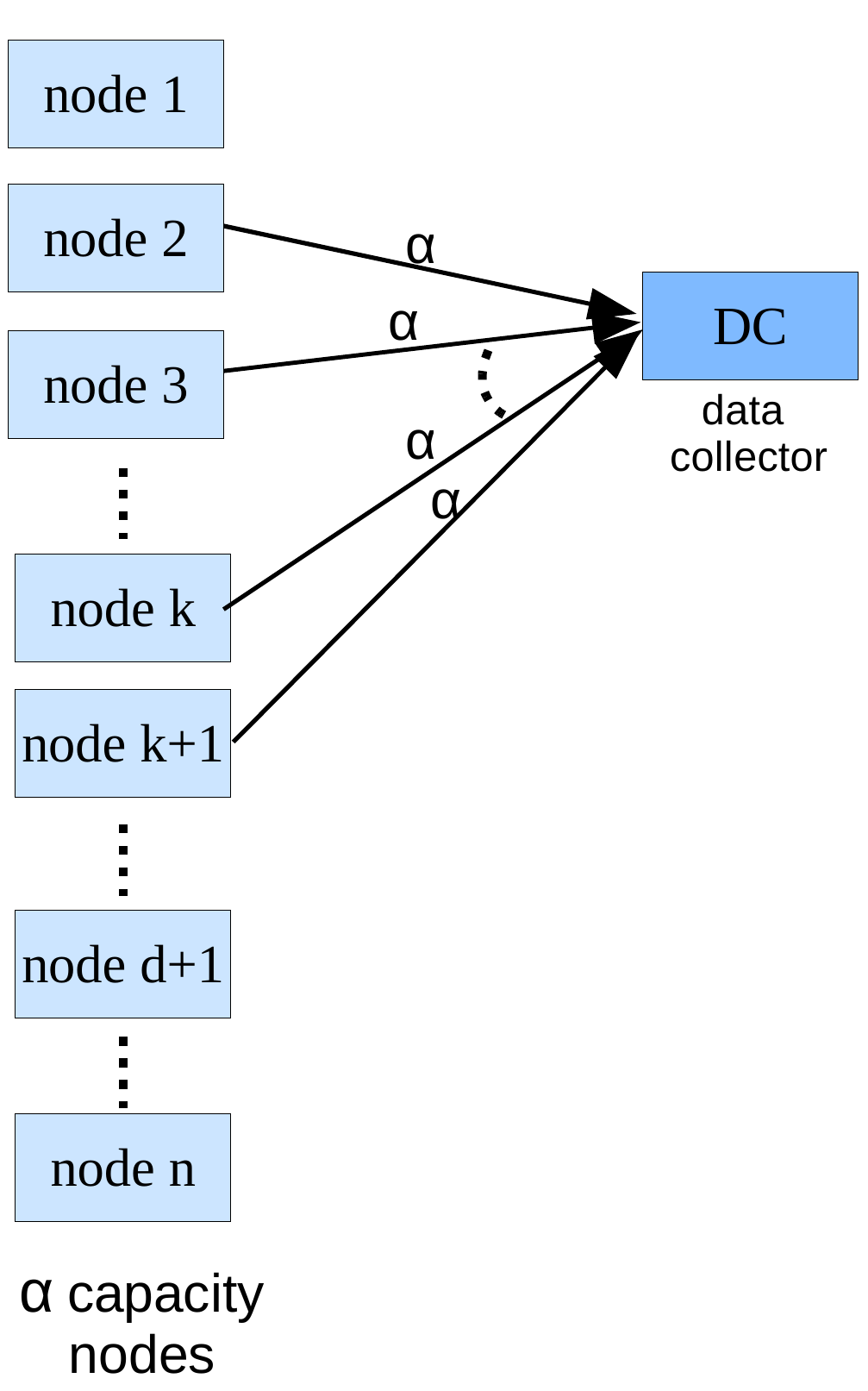}\label{fig:intro_recon}}
\hspace{.1\textwidth}
\subfloat[]{\includegraphics[trim=0in 1.81in 7in 0in, clip, width=0.3\textwidth]{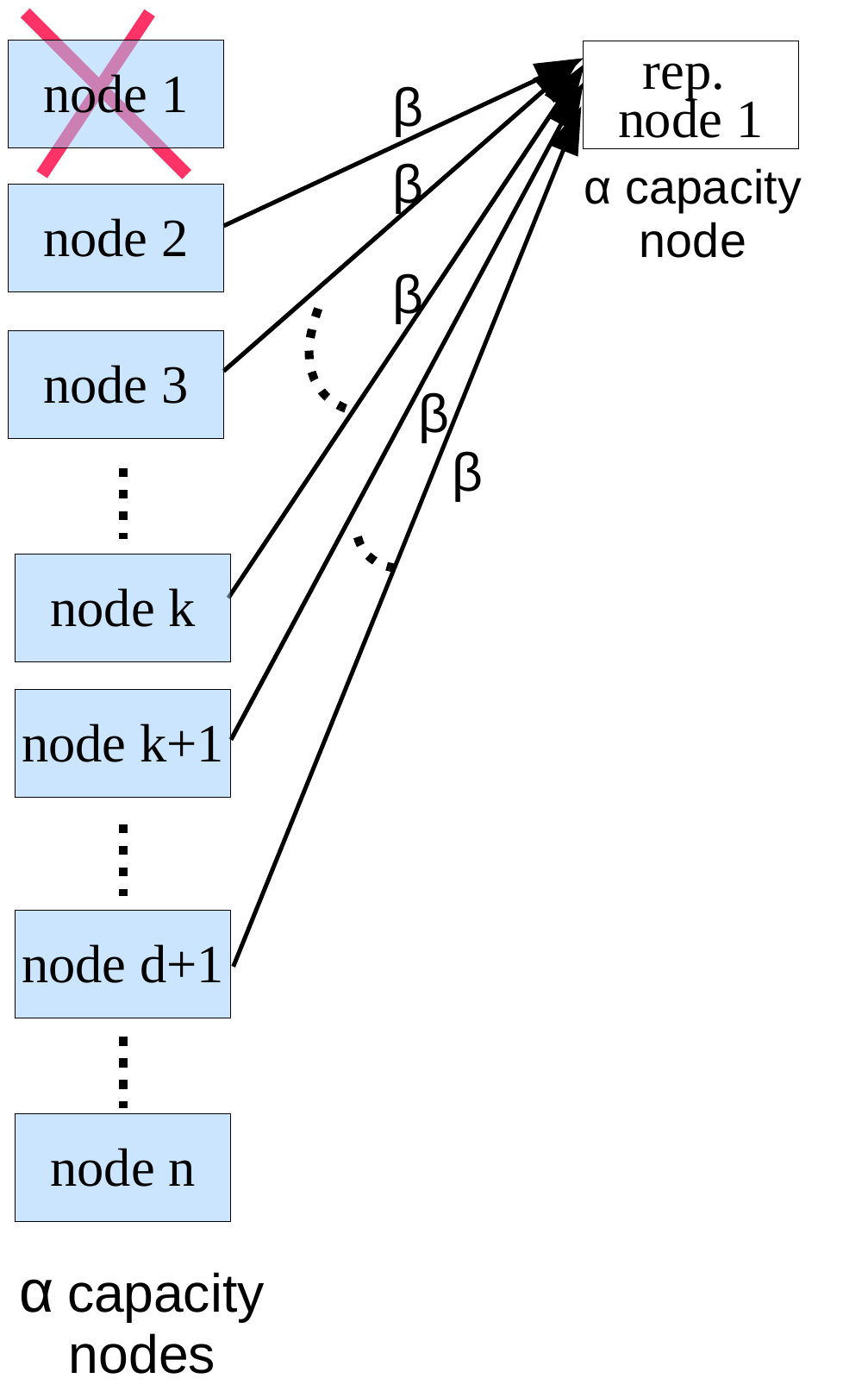}\label{fig:intro_regen}}
\caption{\small The regenerating codes setup: (a) data reconstruction, and (b) repair of a failed node.}
\label{fig:completegraph}
\end{figure}

Apart from this new parameter $\alpha$, two other parameters $(d, \beta)$ are associated with regenerating codes.  Thus we have
\[
\{ q, \ [n, \ k, \ d], \  (\beta, \ \alpha, B)\}
\]
as the parameter set of a regenerating code. Under the definition of regenerating codes introduced in \cite{DimKan1}, a failed node is permitted to connect to an arbitrary subset of $d$ nodes out of the remaining $(n-1)$ nodes while downloading $\beta \leq \alpha$ symbols from each node. The total amount  $d\beta$ of data downloaded for repair purposes is termed the \textit{repair bandwidth}. Typically, with a regenerating code, the average repair bandwidth $d\beta$ is small compared to the size of the file $B$. Fig.~\ref{fig:intro_recon} and Fig.~\ref{fig:intro_regen} illustrate reconstruction and node repair respectively, also depicting the relevant parameters.

The cut-set bound of network coding can be invoked to show that the parameters of a regenerating code must necessarily satisfy \cite{YunDimKan}:
\bea B & \leq &
\sum_{i=0}^{k-1} \min\{\alpha,(d-i)\beta\}. \label{eq:cut_set_bound} \eea It is desirable to minimize both $\alpha$ as well as $\beta$ since  minimizing $\alpha$ results in a minimum storage solution while minimizing $\beta$~(for a fixed $d$) results in a solution that minimizes the repair bandwidth.  It turns out that there is a tradeoff between $\alpha$ and $\beta$.  The two extreme points in this tradeoff are termed the minimum storage regenerating~(MSR) and minimum bandwidth regenerating~(MBR) points respectively.  The parameters $\alpha$ and $\beta$ for the MSR point on the tradeoff can be obtained by first minimizing $\alpha$ and then minimizing $\beta$ to obtain
\bea
\alpha_{\text{MSR}} & = & \frac{B}{k} , \nonumber \\
\beta_{\text{MSR}} & = & \frac{B}{k(d-k+1)}. \label{eq:MSR_parameters} \eea
Reversing the order, leads to the MBR point which thus corresponds to
\bea
\beta_{\text{MBR}} & = & \frac{2B}{k(2d-k+1)} , \nonumber \\
\alpha_{\text{MBR}} & = & \frac{2dB}{k(2d-k+1)} .
 \label{eq:MBR_parameters} \eea

The focus of the present paper is on the MSR point.  Note that regenerating codes with $(\alpha = \alpha_{\text{MSR}})$ and $(\beta = \beta_{\text{MSR}})$ are necessarily MDS codes over the vector alphabet $\mathbb{F}_q^{\alpha}$.  This follows since the ability to reconstruct the data from any arbitrary $k$ nodes necessarily implies a minimum distance $d_{\min}=n-k+1$.  Since the code size equals $ \left( q^{\alpha} \right) ^k$, this meets the Singleton bound causing the code to be an MDS code.

\subsection{Choice of the Parameter $\beta$}\label{subsec:beta_1}
Let us next rewrite \eqref{eq:MSR_parameters} in the form
\bea
\alpha_{\text{MSR}} & = & \beta_{\text{MSR}} (d-k+1)  \nonumber \\
B & = & \beta_{\text{MSR}} (d-k+1)(k).  \label{eq:beta_as_quantum}\eea
Thus if one is able to construct an $[n,\;k,\;d]$ MSR code with repair bandwidth achieving the cut-set bound for a given value of $\beta$, then both $\alpha_{\text{MSR}}=(d-k+1) \beta_{\text{MSR}}$ and the size $B=k \, \alpha_{\text{MSR}}$ of the file are necessarily fixed.   It thus makes sense to speak of an achievable triple
\[
(\beta, \ \ \alpha=(d-k+1) \beta, \ \ B= k \alpha).
\]
However if a triple $(\beta, \alpha, B)$ is achievable, then so is the triple $(\ell \beta, \ell \alpha, \ell B)$ simply through a process of divide and conquer, i.e., we divide up the message file into $\ell$ sub-files and apply the code for $(\beta, \alpha, B)$ to each of the $\ell$ sub-files.  Hence, codes that are applicable for the case $\beta=1$, are of particular importance as they permit codes to be constructed for every larger integral value of $\beta$. In addition, a code with small $\beta$ will involve manipulating a smaller number of message symbols and hence will in general, be of lesser complexity.  For these reasons, in the present paper, codes are constructed for the case $\beta=1$. Setting $\beta=1$ at the MSR point yields
\beq \alpha_{\text{MSR}}=d-k+1 . \label{eq:MSR_beta1_parameters}\eeq
Note that when $\alpha=1$, we have $B=k$ and meeting the cut-set bound would imply $d = k$. In this case, any $[n,k]$-MDS code will achieve the bound. Hence, we will consider $\alpha > 1$ throughout.

\subsection{Additional Terminology}
\subsubsection{Exact versus Functional Repair}
In general, the cut-set bound~(as derived in~\cite{DimKan1}) applies to functional-repair, that is, it applies to networks which replace a failed node with a replacement node which can carry out all the functions of the earlier failed node, but which does not necessarily store the same data. Thus, under functional-repair, there is need for the network to inform all nodes in the network of the replacement. This requirement is obviated under exact-repair, where a replacement node stores exactly the same data as was stored in the failed node. We will use the term {\em exact-repair MSR code} to denote a regenerating code operating at the minimum storage point, that is capable of exact-repair.

\subsubsection{Systematic Codes}
A systematic regenerating code can be defined as a regenerating code designed in such a way that the $B$ message symbols are explicitly present amongst the $k \alpha$ code symbols stored in a select set of $k$ nodes, termed as the systematic nodes. Clearly, in the case of systematic regenerating codes, exact-repair of the systematic nodes is mandated. A data-collector connecting to the $k$ systematic nodes obtains the $B$ message symbols in an uncoded form, making systematic nodes a preferred choice for data recovery. This makes the fast repair of systematic nodes a priority, motivating the interest in  minimizing the repair bandwidth for the exact-repair of systematic nodes.

~

The immediate question that this raises, is as to whether or not the combination of (a) restriction to repair of systematic nodes and (b) requirement for exact-repair of the systematic nodes leads to a bound on the parameters $(\alpha, \beta)$ different from the cut-set bound.  It turns out that the same bound on the parameters $(\alpha, \beta)$ appearing in \eqref{eq:MSR_parameters} still applies and this is established in Section~\ref{sec:notation}.

\subsection{Exact-repair MSR Codes as Network Codes}\label{subsec:net_cod}
The existence of regenerating codes for the case of functional-repair was proved~(\cite{DimKan1,YunDimKan}) after casting the reconstruction and repair problems as a multicast network coding problem, and using random network codes to achieve the cut-set bound. As shown in our previous work \cite{ourNCC_NC}, construction of exact-repair MSR codes for the repair of systematic nodes is most naturally mapped to a non-multicast problem in network coding, for which very few results are available.

\begin{figure}[h]
\centering
\includegraphics[trim=8in 3.3in 9in 1.5in, clip=true, width=0.8\textwidth]{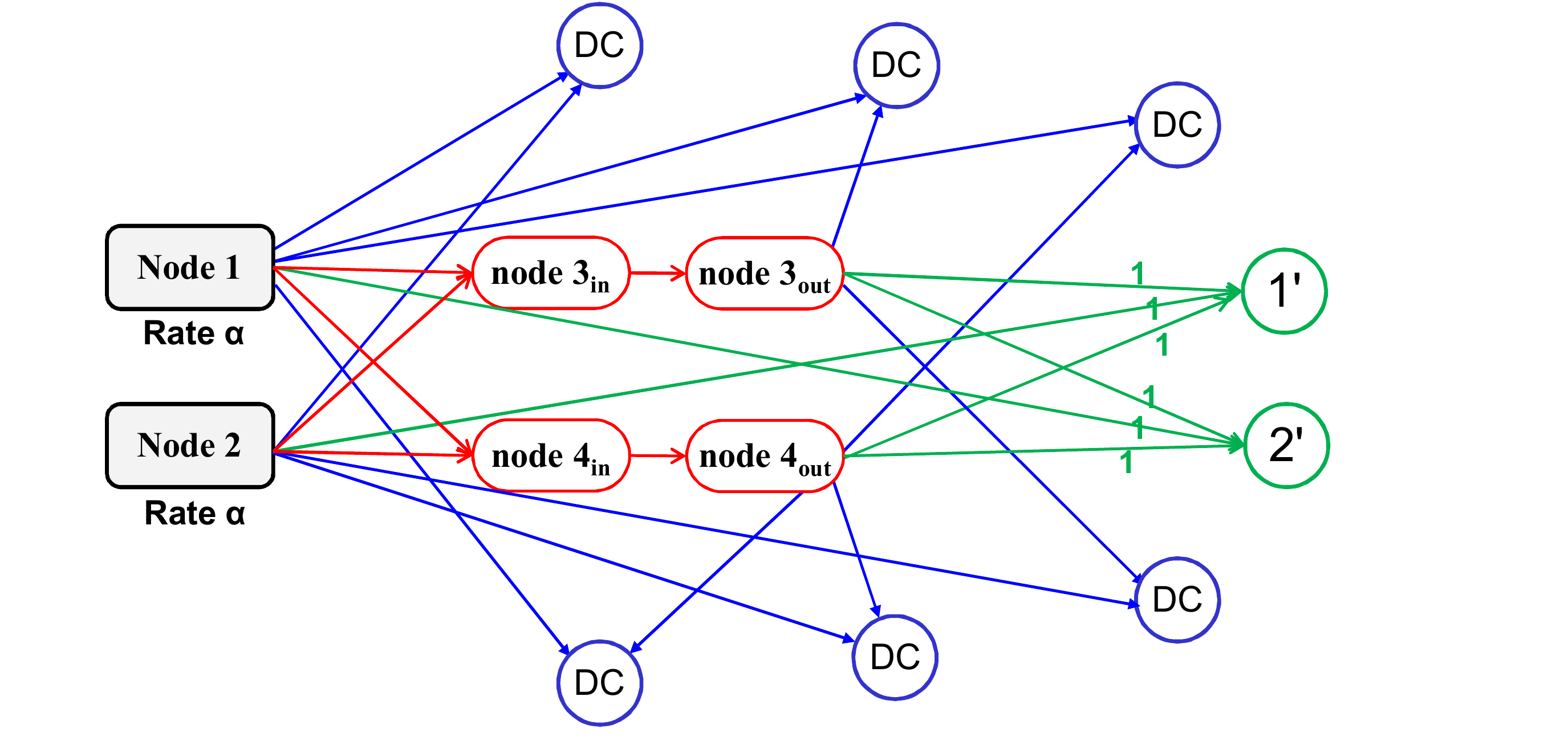}
\caption{\small The MSR code design problem for the exact-repair of just the systematic nodes, as a non-multicast network coding problem. Here, $[n=4, \ k=2 \ d=3]$ with $\beta=1$ giving $(\alpha=2, \ B=4)$. Unmarked edges have capacity $\alpha$. Nodes labelled \textit{DC} are data-collector sinks, and those labelled \textit{$l'$} are replacement node sinks.} \label{fig:stoMult423}
\end{figure}

The non-multicast network for the parameter set $[n=4,\ k=2,\ d=3]$ with $\beta=1$ is shown in Fig.~\ref{fig:stoMult423}. In general, the network can be viewed as having $k$ source nodes, corresponding to the $k$ systematic nodes, generating $\alpha$ symbols each per channel use. The parity nodes correspond to downlink nodes in the graph. To capture the fact that a parity node can store only $\alpha$ symbols, it is split~(as in~\cite{YunDimKan}) into two parts connected by a link of capacity $\alpha$ : parity node $m$ is split into $m_{\text{in}}$ and $m_{\text{out}}$ with all incoming edges arriving at $m_{\text{in}}$ and all outgoing edges emanating from $m_{\text{out}}$.

The sinks in the network are of two types. The first type correspond to data-collectors which connect to an arbitrary collection of $k$ nodes in the network for the purposes of data reconstruction. Hence there are ${n}\choose{k}$ sinks of this type. The second type of sinks represent a replacement node that is attempting to duplicate a failed systematic node, with the node replacing systematic node $\ell$ denoted by $\ell'$. Sinks of this type connect to an arbitrary set of $d$ out of the remaining $(n-1)$ nodes, and hence they are $k { {n-1}\choose{d}}$ in number. It is the presence of these sinks that gives the problem a non-multicast nature.

Thus, the present paper provides an instance where explicit code constructions achieve the cut-set bound for a non-multicast network, by exploiting the specific structure of the network.

~

\paragraph*{Relation Between $\beta$ and Scalar/Vector Network Coding}
The choice of $\beta$ as unity~(as in Fig.~\ref{fig:stoMult423}) may be viewed as an instance of scalar network coding. Upon increase in the value of $\beta$, the capacity of each data pipe is increased by a factor of $\beta$, thereby transforming the problem into a \textit{vector network coding} problem. Thus, $\beta=1$ implies
the absence of \textit{symbol extension}, which in general, reduces the complexity of system implementation and is thus of greater practical interest.

\subsection{Results of the Present Paper}
The primary results of the present paper are:
\bit
\item The construction of a family of MDS codes for $d = n-1 \geq 2k-1$ that enable exact-repair of systematic nodes while achieving the cut-set bound on repair bandwidth. We have termed this code the MISER~\footnote{Short for an MDS, Interference-aligning, Systematic, Exact-Regenerating code, that is miserly in terms of bandwidth expended to repair a systematic node.} code.
\item Proof that interference alignment is \textit{necessary} for every exact-repair MSR code.
\item The proof  of non-existence of linear exact-repair MSR codes for $d < 2k-3$ in the absence of symbol extension~(i.e., $\beta=1$).  This result is clearly of interest in the light of on-going efforts to construct exact-repair codes with $\beta=1$ meeting the cut-set bound~\cite{WuDimISIT,ourAllerton,ourITW,DimSearch,WuArxiv,ourInterior_pts,Changho,ourProductMatrix,puyol}.
\item The construction, also explicit, of an MSR code for $d=k+1$. For most values of the parameters, $d=k+1$ falls under the $d<2k-3$ regime, and in light of the non-existence result above, exact-repair is not possible. The construction does the next best thing, namely, it carries out repair that is approximately exact~\footnote{The code consists of an exact-repair part along with an auxiliary part whose repair is not guaranteed to be exact. This is explained in greater detail in Section~\ref{sec:MDSplus}.}.
\eit
~

Note that the only explicit codes of the MDS type to previously have been constructed are for small values of parameters, $[n=4, \ k=2,\ d=3]$ and $[n=5, \ k=3,\ d=4]$.  Prior work is described in greater detail in Section~\ref{sec:priorWork}.

~

The remainder of the paper is organized as follows. A brief overview of the prior literature in this field is given in the next section, Section~\ref{sec:priorWork}. The setting and notation are explained in Section~\ref{sec:notation}.  The appearance of interference alignment in the context of distributed storage for construction of regenerating codes is detailed in Section~\ref{sec:intf_align} along with an illustrative example. Section~\ref{sec:gen_explicit} describes the MISER code. The non-existence of linear exact-repair MSR codes for $d < 2k-3$ in the absence of symbol extension can be found in Section~\ref{sec:non_exist_alpha_3}, along with the proof establishing the necessity of interference alignment. Section~\ref{sec:MDSplus} describes the explicit construction of an MSR code for $d=k+1$.  The final section, Section~\ref{sec:conclusion}, draws conclusions.

\section{Prior Work}\label{sec:priorWork}
The concept of regenerating codes, introduced in~\cite{DimKan1,YunDimKan}, permit storage nodes to store more than the minimal $B/k$ units of data in order to reduce the repair bandwidth. Several distributed systems are analyzed, and estimates of the mean node availability in such systems are obtained. Using these values, the substantial performance gains offered by regenerating codes in terms of bandwidth savings are demonstrated.

The problem of minimizing repair bandwidth for the \textit{functional} repair of nodes is considered in~\cite{DimKan1,YunDimKan} where it is formulated as a multicast network-coding problem in a network having an infinite number of nodes. A cut-set lower bound on the repair bandwidth is derived. Coding schemes achieving this bound are presented in~\cite{YunDimKan, WuAchievable} which however, are non-explicit.  These schemes require large field size and the repair and reconstruction algorithms are also of high complexity.

Computational complexity is identified as a principal concern in the practical implementation of distributed storage codes in~\cite{Complexi} and a treatment of the use of random, linear, regenerating codes for achieving functional-repair can be found there.

The authors in~\cite{WuDimISIT} and~\cite{ourAllerton} independently introduce the notion of exact-repair.  The idea of using interference alignment in the context of exact-repair codes for distributed storage appears first in~\cite{WuDimISIT}. Code constructions of the MDS type are provided, which meet the cut-set lower bound when $k=2$. Even here, the constructions are not explicit, and have large complexity and field-size requirement.

The first explicit construction of regenerating codes for the MBR point appears in \cite{ourAllerton}, for the case $d=n-1$.  These codes carry out uncoded exact-repair and hence have zero repair complexity. The required field size is of the order of $n^2$, and in terms of minimizing bandwidth,  the codes achieve the cut-set bound.

A computer search for exact-repair MSR codes for the parameter set $[n=5,~k=3,~d=4], \ ~\beta=1$, is carried out in~\cite{DimSearch}, and for this set of parameters, codes for several values of field size are obtained.

A slightly different setting, from the exact-repair situation  is considered in~\cite{WuArxiv}, where optimal MDS codes are given for the parameters $d=k+1$ and $n>2k$. Again, the schemes given here are non-explicit, and have high complexity and large field-size requirement.

We next describe the setting and notation to be used in the current paper.

\section{Setting and Notation} \label{sec:notation}
The distributed storage system considered in this paper consists of $n$ storage nodes, each having the capacity to store $\alpha$ symbols. Let $\underline{\mathbf{u}}$ be the message vector of length $B$ comprising of the $B$ message symbols. Each message symbol can independently take values from $\mathbb{F}_q$, a finite field of size $q$.

In this paper, we consider only linear storage codes. As in traditional coding theory, by a linear storage code, we mean that every stored symbol is a linear combination of the message symbols, and only linear operations are permitted on the stored symbols. Thus all symbols considered belong to $\mathbb{F}_q$.

For $m=1,\ldots,n$, let the $(B \times \alpha)$ matrix $\mathbf{G}^{(m)}$ denote the generator matrix of node $m$. Node $m$ stores the following $\alpha$ symbols
\beq \underline{\mathbf{u}}^t\mathbf{G}^{(m)}.  \eeq
\noindent
In the terminology of network coding, each column of the nodal generator matrix $\mathbf{G}^{(m)}$ corresponds to the \textit{global kernel}~(linear combination vector) associated to a symbol stored in the node. The $(B \times n \alpha)$ generator matrix for the entire distributed-storage code, is given by
\beq \mathbb{G} \ = \  \begin{bmatrix}
     \mathbf{G}^{(1)} & \mathbf{G}^{(2)} & \cdots & \mathbf{G}^{(n)}
     \end{bmatrix}.
\eeq
Note that under exact-repair, the generator matrix of the code remains unchanged.

We will interchangeably speak of a node as either storing $\alpha$ symbols, by which we will mean the symbols $\underline{\mathbf{u}}^t\mathbf{G}^{(m)}$ or else as storing $\alpha$ vectors, by which we will mean the corresponding set of $\alpha$ global kernels that form the columns of nodal generator matrix $\mathbf{G}^{(m)}$.

We partition the $B(=k\alpha)$-length vector $\underline{\mathbf{u}}$ into $k$ components, $\underline{u}_i$ for $i=1,\ldots,k$, each comprising of $\alpha$ distinct message symbols:
\beq \underline{\mathbf{u}}= \begin{bmatrix} \underline{u}_1 \\ \vdots \\ \underline{u}_k\end{bmatrix}. \eeq
We also partition the nodal generator matrices analogously into $k$ sub-matrices as
\beq \mathbf{G}^{(m)} = \begin{bmatrix} G^{(m)}_1 \vspace{5pt} \\ \vdots \vspace{5pt} \\ G^{(m)}_k \vspace{5pt} \end{bmatrix} \label{eq:notation_1}, \eeq
\noindent
where each $G^{(m)}_i$ is an $(\alpha \times \alpha)$ matrix. We will refer to $G^{(m)}_i$ as the $i^{\text{th}}$ component of $\mathbf{G}^{(m)}$. Thus, node $m$ stores the $\alpha$ symbols
\beq \underline{\mathbf{u}}^t \mathbf{G}^{(m)}  = \sum_{i=1}^{k} \underline{u}^t_i G^{(m)}_i \label{eq:notation_2}. \eeq

~

Out of the $n$ nodes, the first $k$ nodes~(i.e., nodes $1,\ldots,k$) are systematic. Thus, for systematic node $\ell$
\bea
G^{(\ell)}_i =
\left \lbrace \begin{array}{ll}
         I_{\alpha}  &\text{if } i=\ell \\
0_\alpha &\text{if } i\neq \ell
        \end{array} \right. \quad \forall i \in \{1,\ldots,k \},
\eea
where $0_\alpha$ and $I_\alpha$ denote the $(\alpha \times \alpha)$ zero matrix and identity matrix respectively; systematic node $\ell$ thus stores the $\alpha$ message symbols that $\underline{u}_\ell$ is comprised of.

Upon failure of a node, the replacement node connects to an arbitrary set of $d$ remaining nodes, termed as \textit{helper nodes}, downloading $\beta$ symbols from each. Thus, each helper node passes a collection of $\beta$ linear combinations of the symbols stored within the node. As described in Section~\ref{subsec:beta_1}, an MSR code with $\beta=1$ can be used to construct an MSR code for every higher integral value of $\beta$. Thus it suffices to provide constructions for  $\beta=1$ and that is what we do here. When $\beta=1$, each helper node passes just a single symbol. Again, we will often describe the symbol passed by a helper node in terms of its associated global kernel, and hence will often speak of a helper node passing a \textit{vector}~\footnote{A simple extension to the case of $\beta > 1$ lets us treat the global kernels of the $\beta$ symbols passed by a helper node as a \textit{subspace} of dimension at most $\beta$. This `subspace' viewpoint has been found useful in proving certain general results at the MBR point in \cite{ourAllerton}, and for the interior points of the tradeoff in~\cite{ourInterior_pts}.}.

~

Throughout the paper, we use superscripts to refer to node indices, and subscripts to index the elements of a matrix. The letters $m$ and $\ell$ are reserved for node indices; in particular, the letter $\ell$ is used to index systematic nodes. All vectors are assumed to be column vectors.  The vector $\underline{e}_i$ represents the standard basis vector of length $\alpha$, i.e., $\underline{e}_i$ is an $\alpha$-length unit vector with $1$ in the $i$th position and $0$s elsewhere. For a positive integer $p$, we denote the $(p \times p)$ zero matrix and the $(p \times p)$ identity matrix by $0_p$ and $I_p$ respectively.  We say that a set of vectors is \textit{aligned} if the vector-space spanned by them has dimension at most one.

~

We next turn our attention to the question as to whether or not the combination of (a) restriction to systematic-node repair and (b) requirement of exact-repair of the systematic nodes leads to a bound on the parameters $(\alpha, \beta)$ different from the cut-set bound appearing in~\eqref{eq:cut_set_bound}.

The theorem below shows that the cut-set bound comes into play even if functional repair of a single node is required.

\begin{thm}
Any $[n, \ k, \ d]$-MDS regenerating code~(i.e., a regenerating code satisfying $B=k\alpha$) that guarantees the functional-repair of even a single node,
must satisfy the cut-set lower bound of~\eqref{eq:cut_set_bound} on repair bandwidth, i.e., must satisfy
\beq \beta \geq \frac{B}{k(d-k+1)}. \eeq
\end{thm}

\begin{IEEEproof}
First, consider the case when $\beta=1$. Let $\ell$ denote the node that needs to be repaired, and let $\{m_i \mid i=1, \ldots, d\}$ denote the $d$ helper nodes assisting in the repair of node $\ell$. Further, let $\{\underline{\mathbf{\gamma}}^{(m_i, \; \ell)}\mid i=1,\ldots,d\}$ denote the vectors passed by these helper nodes. At the end of the repair process, let the $(B \times \alpha)$ matrix $\mathbf{G}^{(\ell)}$ denote the generator matrix of the replacement node~(since we consider only functional-repair in this theorem, $\mathbf{G}^{(\ell)}$ need not be identical to the generator matrix of the failed node).

Looking back at the repair process, the replacement node obtains $\mathbf{G}^{(\ell)}$ by operating linearly on the collection of $d$ vectors $\{\underline{\mathbf{\gamma}}^{(m_i, \; \ell)}\mid i=1,\ldots,d\}$ of length $B$.  This, in turn, implies that the dimension of the nullspace of the matrix
\beq \begin{bmatrix} \mathbf{G}^{(\ell)} & \underline{\mathbf{\gamma}}^{(m_1,\; \ell)} & \cdots & \underline{\mathbf{\gamma}}^{(m_d,\; \ell)} \end{bmatrix} \label{eq:nullspace_alpha}\eeq
should be greater than or equal to the dimension of $\mathbf{G}^{(l)}$, which is $\alpha$. However, the MDS property requires that at the end of the repair process, the global kernels associated to any $k$ nodes be linearly independent, and in particular, that the matrix
\beq \begin{bmatrix}\mathbf{G}^{(\ell)} & \underline{\mathbf{\gamma}}^{(m_1,\; \ell)} & \cdots & \underline{\mathbf{\gamma}}^{(m_{k-1},\; \ell)} \end{bmatrix} \eeq have full-rank. It follows that we must have
\[
d \ \geq \ k-1+\alpha.
\]

The proof for the case $\beta>1$, when every helper node passes a set of $\beta$ vectors, is a straightforward extension that leads to:
\beq d\beta \ \geq\ (k-1)\beta + \alpha. \eeq
Rearranging the terms in the equation above, and substituting $\alpha = \frac{B}{k}$ leads to the desired result.
\end{IEEEproof}

~

\noindent
Thus, we recover equation~\eqref{eq:MSR_parameters}, and in an optimal code with $\beta=1$, we will continue to have
\[
d \ = \ k-1+\alpha.
\]
In this way, we have shown that even in the setting that we address here, namely that of the exact-repair of the systematic nodes leads us to the same cut-set bound on repair bandwidth as in ~\eqref{eq:cut_set_bound}.
\noindent
The next section explains how the concept of interference alignment arises in the distributed-storage context.

\section{Interference Alignment in Regenerating Codes}\label{sec:intf_align}
The idea of interference alignment has recently been proposed in \cite{CadJafar}, \cite{MotKhan}  in the context of wireless communication.  The idea here is to design the signals of multiple users in such a way that at every receiver, signals from all the unintended users occupy a subspace of the given space, leaving the remainder of the space free for the signal of the intended user.

In the distributed-storage context, the concept of `interference' comes into play during the exact-repair of a failed node in an MSR code. We present the example of a systematic MSR code with $[n=4, \; k=2, \; d=3]$ and $\beta=1$, which gives $(\alpha=d-k+1=2,\; B=k\alpha = 4)$. Let $\{ u_1, \ u_2, \ u_3, \ u_4 \}$ denote the four message symbols.  Since $k=2$ here, we may assume that nodes $1$ and $2$ are systematic and that node $1$ stores $\{ u_1, \ u_2\}$ and node $2$ stores $\{ u_3, \ u_4 \}$. Nodes $3$ and $4$ are then the parity nodes, each storing two linear functions of the message symbols.

\begin{figure}[h]
\centering
\includegraphics[trim= 0.1in 6.4in 4in 0in, clip=true,width=\textwidth]{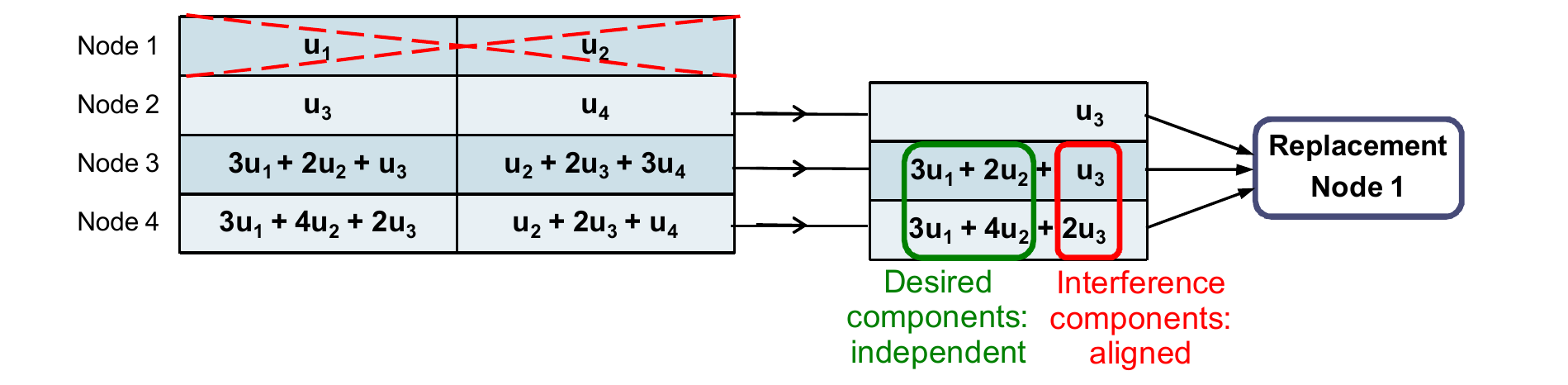} 
\caption{\small Illustration of interference alignment during exact-repair of systematic node $1$.}
\label{fig:fig_42msr_regensys}
\end{figure}

Consider repair of systematic node $1$ wherein the $d=3$ nodes, nodes $2$, $3$ and $4$, serve as helper nodes.  The second systematic node, node $2$, can only pass a linear combination of message symbols $u_3$ and $u_4$.  The two symbols passed by the parity nodes are in general, functions of all four message symbols: $(a_1 u_1 + a_2 u_2 + a_3 u_3 + a_4 u_4)$ and  $(b_1 u_1 + b_2 u_2 + b_3 u_3 + b_4 u_4)$ respectively.

Using the symbols passed by the three helper nodes, the replacement of node $1$ needs to be able to recover message symbols $\{u_1,u_2\}$.  For obvious reasons, we will term  $(a_1 u_1 + a_2 u_2 )$ and $(b_1 u_1 + b_2 u_2)$ as the \textit{desired} components of the messages passed by parity nodes $3$ and $4$ and the terms $(a_3 u_3 + a_4 u_4)$ and $(b_3 u_3 + b_4 u_4)$ as \textit{interference} components.

Since node $2$ cannot provide any information pertaining to the desired symbols $\{ u_1, \ u_2\}$, the replacement node must be able to recover the desired symbols from the desired components $(a_1 u_1 + a_2 u_2 )$ and $(b_1 u_1 + b_2 u_2)$ of the messages passed to it by the parity nodes $3$ and $4$. To access the desired components, the replacement node must be in a position to subtract out the interference components $(a_3 u_3 + a_4 u_4)$ and $(b_3 u_3 + b_4 u_4)$ from the received linear combinations $(a_1 u_1 + a_2 u_2 + a_3 u_3 + a_4 u_4)$ and  $(b_1 u_1 + b_2 u_2 + b_3 u_3 + b_4 u_4)$; the only way to subtract out the interference component is by making use of the linear combination of $\{u_3,u_4\}$ passed by node $2$. It follows that this can only happen if the interference components $(a_3 u_3 + a_4 u_4)$ and $(b_3 u_3 + b_4 u_4)$ are aligned, meaning that they are scalar multiples of each other.

An explicit code over $\mathbb{F}_5$ for the parameters chosen in the example is shown in Fig.~\ref{fig:fig_42msr_regensys}. The exact-repair of systematic node $1$ is shown, for which the remaining nodes pass the first of the two symbols stored in them. Observe that under this code, the interference component in the two symbols passed by the parity nodes are aligned in the direction of  $u_3$, i.e., are scalar multiples of $u_3$. Hence node $2$ can simply pass $u_3$ and the replacement node can then make use of $u_3$ to cancel~(i.e., subtract out) the interference.

In the context of regenerating codes, interference alignment was first used by Wu et al. \cite{WuDimISIT} to provide a scheme~(although, not explicit) for the exact-repair at the MSR point. However, interference alignment is employed only to a limited extent as only a portion of the interference components is aligned and as a result, the scheme is optimal only for the case $k=2$.

In the next section, we describe the construction of the MISER code which aligns interference and achieves the cut-set bound on the repair bandwidth for repair of systematic nodes. This is the \textit{first} interference-alignment-based explicit code construction that meets the cut-set bound.

\section{Construction of the MISER Code}\label{sec:gen_explicit}
In this section we provide an explicit construction for a systematic, MDS code that achieves the lower bound on repair bandwidth for the exact-repair of systematic nodes and which we term as the MISER code.   We begin with an illustrative example that explains the key ideas behind the construction.  The general code construction for parameter sets of the form  $n=2k,~d=n-1$ closely follows the construction in the example. A simple, code-shortening technique is then employed to extend this code construction to the more general parameter set $n \geq 2k,~d=n-1$.

The construction technique can also be extended to the even more general case of arbitrary $n$, $d \geq 2k-1$, under the added requirement however, that the replacement node connect to all of the remaining systematic nodes.

\subsection{An Example} \label{sec:example}
The example deals with the parameter set, $[n=6,\;k=3,\;d=5]$, $\beta=1$, so that $(\alpha=d-k+1=3,\;B=k\alpha=9)$.  We select $\mathbb{F}_7$ as the underlying finite field so that all message and code symbols are drawn from $\mathbb{F}_7$.  Note that we have $\alpha=k=3$ here.  This is true in general: whenever $n=2k$ and $d=n-1$, we have $\alpha=d-k+1=k$ which simplifies the task of code construction.

~

\subsubsection{Design of Nodal Generator Matrices}

As $k=3$, the first three nodes are systematic and store data in uncoded form. Hence
\beq
\mathbf{G}^{(1)} = \begin{bmatrix} I_3 \vspace{2pt} \\ 0_3 \vspace{2pt} \\ 0_3 \end{bmatrix} , \
\mathbf{G}^{(2)} = \begin{bmatrix} 0_3 \vspace{2pt} \\ I_3 \vspace{2pt} \\ 0_3 \end{bmatrix} , \
\mathbf{G}^{(3)} = \begin{bmatrix} 0_3 \vspace{2pt} \\ 0_3 \vspace{2pt} \\ I_3 \end{bmatrix}~.
\eeq
A key ingredient of the code construction presented here is the use of a Cauchy matrix~\cite{cauchy}. Let \beq {\Psi}_3 = \left[ \resizebox{!}{!}{\begin{tabular}{*{3}{c}}
${\psi}_1^{(4)}$ & ${\psi}_1^{(5)}$ & ${\psi}_1^{(6)}$ \vspace{2pt} \\
${\psi}_2^{(4)}$ & ${\psi}_2^{(5)}$ & ${\psi}_2^{(6)}$ \vspace{2pt} \\
${\psi}_3^{(4)}$ & ${\psi}_3^{(5)}$ & ${\psi}_3^{(6)}$
\end{tabular}} \right] \label{eq:cauchy} \eeq
be a $(3 \times 3)$ matrix such that each of its sub-matrices is full rank. Cauchy matrices have this property and in our construction, we will assume ${\Psi}_3$ to be a Cauchy matrix.

~

\begin{figure}
 \centering
\includegraphics[trim=0in 0.8in 0 0, clip=true, width=\textwidth]{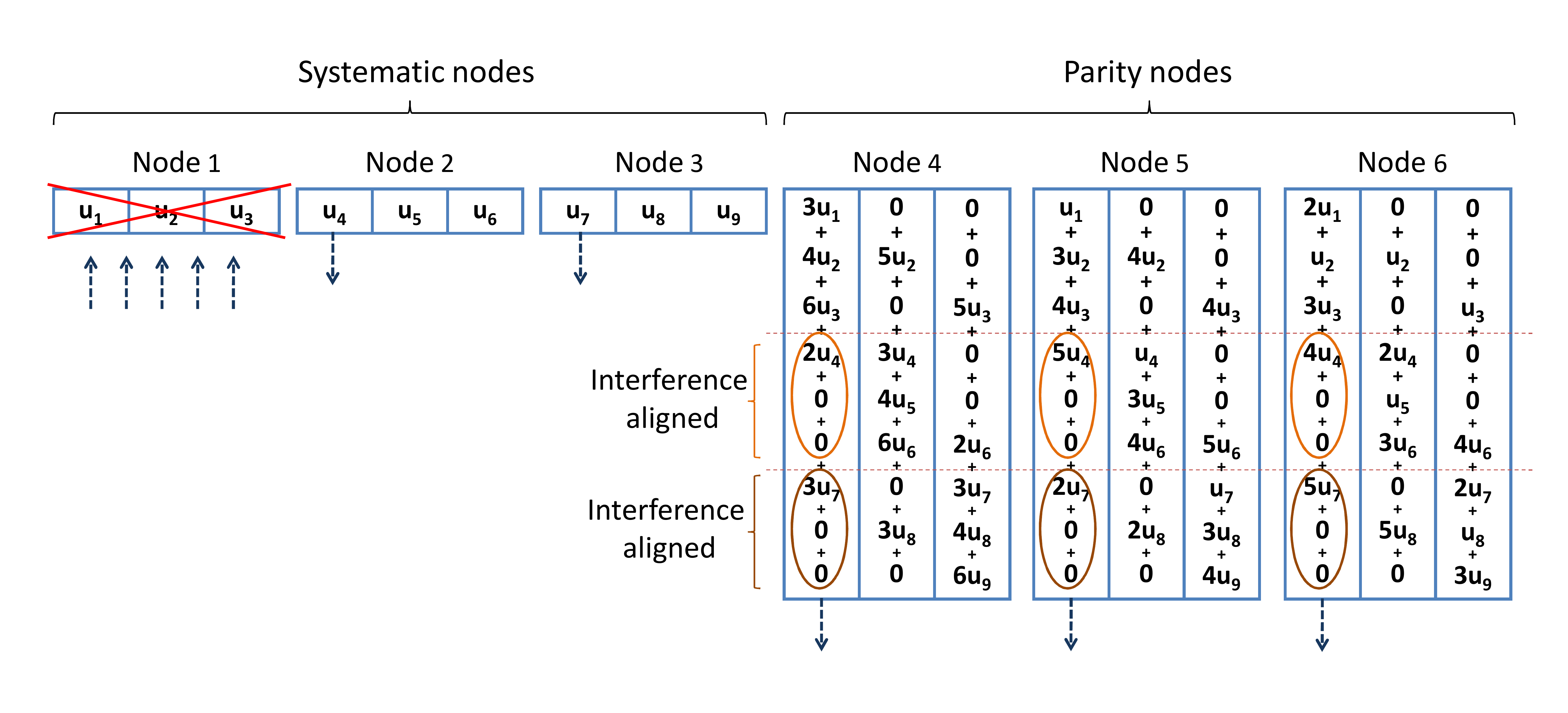}
\caption{\small An example of the $[6, \; 3, \; 5]$ MISER code over $\mathbb{F}_7$. Here, $\{u_1,\ldots,u_9\}$ denote the message symbols and the code symbols stored in each of the nodes are shown. Exact-repair of node $1$ is also depicted.}
\label{fig:example_635}
\end{figure}

We choose the generator matrix of parity node $m~(m=4,5,6)$ to be
\beq \mathbf{G}^{(m)} =  \left[ \resizebox{!}{!}{\renewcommand{\arraystretch}{1.2}\begin{tabular}{*{3}{c}}
$2{\psi}_1^{(m)} $&$ 0 $&$ 0 $  \\
$2{\psi}_2^{(m)} $&$ {\psi}_1^{(m)}  $&$ 0$ \\
$2{\psi}_3^{(m)}  $&$ 0          $&$ {\psi}_1^{(m)}  $  \\ \hline \vspace{-11pt} \\
$ {\psi}_2^{(m)}          $&$ 2{\psi}_1^{(m)} $&$ 0 $  \\

$ 0          $&$ 2{\psi}_2^{(m)} $&$ 0   $  \\
$0$ &$ 2{\psi}_3^{(m)}          $&${\psi}_2^{(m)}$   \\ \hline \vspace{-11pt} \\

$ {\psi}_3^{(m)}  $&$ 0          $&$ 2{\psi}_1^{(m)} $  \\
$ 0          $&$ {\psi}_3^{(m)}  $&$ 2{\psi}_2^{(m)} $  \\
$ 0          $&$ 0          $&$ 2{\psi}_3^{(m)}$  \\
\end{tabular}} \right], \eeq
where the location of the non-zero entries of the $i$th sub-matrix are restricted to lie either along the diagonal or else within the $i$th column.
The generator matrix is designed keeping in mind the need for interference alignment and this will be made clear in the discussion below concerning the exact-repair of systematic nodes. The choice of scalar `$2$' plays an important role in the data reconstruction property; the precise role of this scalar will become clear when this property is discussed. An example of the $[6, \; 3, \; 5]$ MISER code over $\mathbb{F}_7$ is provided in Fig.~\ref{fig:example_635}, where the Cauchy matrix $\Psi$ is chosen as
\beq \Psi = \left[\begin{tabular}{>{$}c<{$}>{$}c<{$}>{$}c<{$}}
             5 & 4 & 1 \\ 2 & 5 & 4 \\ 3 & 2 & 5
            \end{tabular} \right].
\eeq
Also depicted in the figure is the exact-repair of node $1$, for which each of the remaining nodes pass the first symbol that they store. It can be seen that the first symbols stored in the three parity nodes $4$, $5$ and $6$ have their  interference components (components $2$ and $3$) aligned and their desired components (component $1$) linearly independent.

~

The key properties of the MISER code will be established in the next section, namely:
\bit \item that the code is an MDS code over alphabet $\mathbb{F}_q^\alpha$ and this property enables data reconstruction and \item that the code has the ability to carry out exact-repair of the systematic nodes while achieving the cut-set bound on repair bandwidth. \eit

We begin by establishing the exact-repair property.

~

\subsubsection{Exact-repair of Systematic Nodes}

Our algorithm for systematic node repair is simple.  As noted above, each node stores $\alpha=k$ symbols. These $k$ symbols are assumed to be ordered so that we may speak of the first symbol stored by a node, etc. To repair systematic node $\ell$, $1 \leq \ell \leq k$, each of the remaining nodes passes their respective $\ell$th symbol.

Suppose that in our example construction here, node $1$ fails. Each of the parity nodes then pass on their first symbol, or equivalently, in terms of global kernels,  the first column of their generator matrices for the repair of node $1$. Thus, from nodes $4,\ 5,$ and $6$, the replacement node obtains
\beq \hspace{-2pt}
\left[\hspace{-2pt} \resizebox{1.2cm}{!}{\begin{tabular}{c}
         $2{\psi}_1^{(4)}$ \\
     $2{\psi}_2^{(4)}$ \\
     $2{\psi}_3^{(4)}$ \vspace{2pt}\\     \hline
     \vspace{-.3cm} \\
     ${\psi}_2^{(4)}$ \\
     $0$ \\
     $0$ \vspace{2pt}\\     \hline
\vspace{-.3cm} \\
     ${\psi}_3^{(4)}$ \\
     $0$      \\
     $0$
        \end{tabular}} \hspace{-2pt}\right]\hspace{-2pt}, \quad
\left[\hspace{-2pt} \resizebox{1.2cm}{!}{\begin{tabular}{c}
         $2{\psi}_1^{(5)}$ \\
     $2{\psi}_2^{(5)}$ \\
     $2{\psi}_3^{(5)}$ \vspace{2pt}\\     \hline
\vspace{-.3cm} \\
     ${\psi}_2^{(5)}$ \\
     $0$ \\
     $0$ \vspace{2pt}\\     \hline
\vspace{-.3cm} \\
     ${\psi}_3^{(5)}$ \\
     $0$      \\
     $0$
        \end{tabular}} \hspace{-2pt}\right]\hspace{-2pt}, \quad
\left[\hspace{-2pt} \resizebox{1.2cm}{!}{\begin{tabular}{c}
         $2{\psi}_1^{(6)}$ \\
     $2{\psi}_2^{(6)}$ \\
     $2{\psi}_3^{(6)}$ \vspace{2pt}\\     \hline
\vspace{-.3cm} \\
     ${\psi}_2^{(6)}$ \\
     $0$ \\
     $0$ \vspace{2pt}\\     \hline
\vspace{-.3cm} \\
     ${\psi}_3^{(6)}$ \\
     $0$      \\
     $0$
        \end{tabular}} \hspace{-2pt}\right]~.
\eeq

Note that in each of these vectors, the desired~(first) components are a scaled version of the respective columns of the Cauchy matrix $\Psi_3$. The interference~(second and third) components are aligned along the vector $[1 \ \ 0 \  \ 0]^t$.  Thus, each interference component is aligned along a single dimension. Systematic nodes $2$ and $3$ then pass a single vector each that is designed to cancel out this interference. Specifically, nodes $2$ and $3$ respectively pass the vectors
\beq
\left[\hspace{-2pt} \resizebox{0.6cm}{!}{\begin{tabular}{c}
         $0$ \\
     $0$ \\
     $0$ \\     \hline
     $1$ \\
     $0$ \\
     $0$ \\     \hline
     $0$ \\
     $0$      \\
     $0$
        \end{tabular}} \hspace{-2pt}\right], \quad
\left[\hspace{-2pt} \resizebox{0.6cm}{!}{\begin{tabular}{c}
         $0$ \\
     $0$ \\
     $0$ \\     \hline
     $0$ \\
     $0$ \\
     $0$ \\     \hline
     $1$ \\
     $0$      \\
     $0$
        \end{tabular}} \hspace{-2pt}\right]~.
\eeq

The net result is that after interference cancellation has taken place, replacement node $1$ is left with access to the columns of the matrix
\[
\left[ \resizebox{!}{!}{\begin{tabular}{c}
$2{\Psi}_3$   \\ \hline
$0_3$  \\  \hline
$0_3$
\end{tabular}} \right] .
\]
Thus the desired component is a scaled Cauchy matrix ${\Psi}_3$. By multiplying this matrix on the right by $\frac{1}{2}\Psi_3^{-1}$, one recovers
\[
\left[ \resizebox{!}{!}{\begin{tabular}{c}
$I_3$   \\ \hline
$0_3$  \\  \hline
$0_3$
\end{tabular}} \right]
\]
as desired.

Along similar lines, when nodes $2$ or $3$ fail, the parity nodes pass the second or third columns of their generator matrices respectively. The design of generator matrices for the parity nodes is such that interference alignment holds during the repair of either systematic node, hence enabling the exact-repair of all the systematic nodes.

~

\subsubsection{Data Reconstruction~(MDS property)}\label{sec:eg_recon}
For the reconstruction property to be satisfied, a data-collector downloading symbols stored in any three nodes should be able to recover all the nine message symbols. That is, the $(9 \times 9)$ matrix formed by columnwise concatenation of any three nodal generator matrices, should be non-singular. We consider the different possible sets of three nodes that the data-collector can connect to, and provide appropriate decoding algorithms to handle each case.

(a) \textit{Three systematic nodes:} When a data-collector connects to all three systematic nodes, it obtains all the message symbols in uncoded form and hence reconstruction is trivially satisfied.

(b) \textit{Two systematic nodes and one parity node:} Suppose the data-collector connects to systematic nodes $2$ and $3$, and parity node $4$. It obtains all the symbols stored in nodes $2$ and $3$ in uncoded form and proceeds to subtract their effect from the symbols in node $4$. It is thus left to decode the message symbols $\underline{u}_1$, that are encoded using matrix $G^{(4)}_1$ given by
\beq  G^{(4)}_1= \left[ \resizebox{!}{!}{\begin{tabular}{ccc}
$2{\psi}_1^{(4)} $&$0 $&$0 $\\
$ 2{\psi}_2^{(4)} $ &$ {\psi}_2^{(4)} $&$ 0$\\
$ 2{\psi}_3^{(4)} $&$ 0$&$ {\psi}_3^{(4)} $
\end{tabular}} \right]~. \eeq
This lower-triangular matrix is non-singular since by definition, all the entries in a Cauchy matrix are non-zero. The message symbols $\underline{u}_1$ can hence be recovered by inverting $G^{(4)}_1$.

(c) \textit{All three parity nodes:}  We consider next the case when a data-collector connects to all three parity nodes. Let $C_1$ be the $(9 \times 9)$ matrix formed by the columnwise concatenation of the generator matrices of these three nodes.

~

\textit{Claim 1:} The data-collector can recover all the message symbols encoded using the matrix $C_1$, formed by the columnwise concatenation of the generator matrices of the three parity nodes:

\beq C_1 = \left[ \mathbf{G}^{(4)} \quad \mathbf{G}^{(5)} \quad \mathbf{G}^{(6)} \right]. \eeq

\begin{IEEEproof}
We permute the columns of $C_1$ to obtain a second matrix $C_2$ in which the $i^{th}\; (i=1,2,3)$ columns of all the three nodes are adjacent to each other as shown below:
\beq C_2 = \label{eq:invertNonsysStart}
\left[ \resizebox{!}{2.5cm}{\begin{tabular}{ccc|ccc|ccc}
$2{\psi}_1^{(4)} $&$ 2{\psi}_1^{(5)} $&$ 2{\psi}_1^{(6)} $&$ 0  $&$ 0          $&$ 0          $&$ 0  $&$ 0          $&$ 0 $\\
$2{\psi}_2^{(4)}  $&$ 2{\psi}_2^{(5)} $&$ 2{\psi}_2^{(6)}  $&$ {\psi}_1^{(4)}  $&$ {\psi}_1^{(5)}          $&$  {\psi}_1^{(6)}   $&$ 0  $&$ 0          $&$ 0 $\\
$2{\psi}_3^{(4)}  $&$ 2{\psi}_3^{(5)}  $&$ 2{\psi}_3^{(6)} $&$ 0  $&$ 0          $&$ 0          $&$ {\psi}_1^{(4)}  $&$ {\psi}_1^{(5)}          $&$ {\psi}_1^{(6)} $ \vspace{2pt}\\ \hline
\vspace{-.04cm}&\vspace{-.04cm}&\vspace{-.04cm}&\vspace{-.04cm}&\vspace{-.04cm}&\vspace{-.04cm}&\vspace{-.04cm}&\vspace{-.04cm}&\\
${\psi}_2^{(4)}          $&$ {\psi}_2^{(5)}   $&$ {\psi}_2^{(6)}          $&$ 2{\psi}_1^{(4)} $&$ 2{\psi}_1^{(5)} $&$ 2{\psi}_1^{(6)} $&$ 0          $&$ 0  $&$ 0$  \\
$0          $&$ 0  $&$ 0          $&$ 2{\psi}_2^{(4)} $&$ 2{\psi}_2^{(5)} $&$ 2{\psi}_2^{(6)} $&$ 0          $&$ 0  $&$ 0$  \\

$0          $&$ 0   $&$ 0          $&$ 2{\psi}_3^{(4)} $&$ 2{\psi}_3^{(5)} $&$ 2{\psi}_3^{(6)} $&$   {\psi}_2^{(4)}  $&$ {\psi}_2^{(5)}  $&$ {\psi}_2^{(6)}$  \vspace{2pt} \\ \hline
\vspace{-.04cm}&\vspace{-.04cm}&\vspace{-.04cm}&\vspace{-.04cm}&\vspace{-.04cm}&\vspace{-.04cm}&\vspace{-.04cm}&\vspace{-.04cm}&\\
${\psi}_3^{(4)}          $&$ {\psi}_3^{(5)}          $&$ {\psi}_3^{(6)}   $&$ 0          $&$ 0          $&$ 0  $&$ 2{\psi}_1^{(4)} $&$ 2{\psi}_1^{(5)} $&$ 2{\psi}_1^{(6)}$ \\
$0          $&$ 0          $&$ 0  $&$ {\psi}_3^{(4)}          $&$ {\psi}_3^{(5)}            $&$ {\psi}_3^{(6)}  $&$ 2{\psi}_2^{(4)} $&$ 2{\psi}_2^{(5)} $&$ 2{\psi}_2^{(6)}$ \\
$0          $&$ 0          $&$ 0  $&$ 0          $&$ 0          $&$ 0  $&$ 2{\psi}_3^{(4)} $&$ 2{\psi}_3^{(5)} $&$ 2{\psi}_3^{(6)}$ \\
\multicolumn{3}{c}{$\underbrace{\qquad\qquad\qquad\qquad}$}&\multicolumn{3}{c}{$\underbrace{\qquad\qquad\qquad\qquad}$}&\multicolumn{3}{c}{$\underbrace{\qquad\qquad\qquad\qquad}$}\\
\multicolumn{3}{c}{\small{group 1}}&\multicolumn{3}{c}\small{group 2}\hspace{1.3cm}&\multicolumn{3}{c}{\small{group 3}}
 \vspace{-.8cm} \end{tabular}} \right] \nonumber~.\eeq \vspace{.6cm}\\
~

Note that a permutation of the columns does not alter the information available to the data-collector and hence is a permissible operation.  This rearrangement of coded symbols, while not essential, simplifies the proof.  We then post-multiply by a block-diagonal matrix ${\Psi}_3^{-1}$ to obtain the matrix $C_3$ given by
\bea C_3 &=& C_2 \left[ \resizebox{!}{!}{\begin{tabular}{ccc}
          ${\Psi}_3^{-1} $&$ 0_3 $&$ 0_3 $\\
$0_3 $&$ {\Psi}_3^{-1} $&$ 0_3 $\\
$0_3 $&$ 0_3 $&$ {\Psi}_3^{-1} $
\end{tabular}} \right]   \\
&=& \left[ \begin{tabular}{ccc|ccc|ccc}
$2  $&$ 0  $&$ 0  $&$ 0  $&$ 0  $&$ 0  $&$ 0  $&$ 0 $&$ 0  $\\
$0  $&$ 2  $&$ 0  $&$ 1  $&$ 0  $&$ 0  $&$ 0  $&$ 0 $&$ 0  $\\
$0  $&$ 0  $&$ 2  $&$ 0  $&$ 0  $&$ 0  $&$ 1  $&$ 0 $&$ 0  $\\
\hline
$0  $&$ 1  $&$ 0  $&$ 2  $&$ 0  $&$ 0  $&$ 0  $&$ 0 $&$ 0  $\\
$0  $&$ 0  $&$ 0  $&$ 0  $&$ 2  $&$ 0  $&$ 0  $&$ 0 $&$ 0  $\\
$0  $&$ 0  $&$ 0  $&$ 0  $&$ 0  $&$ 2  $&$ 0  $&$ 1 $&$ 0  $\\
\hline
$0  $&$ 0  $&$ 1  $&$ 0  $&$ 0  $&$ 0  $&$ 2  $&$ 0 $&$ 0  $\\
$0  $&$ 0  $&$ 0  $&$ 0  $&$ 0  $&$ 1  $&$ 0  $&$ 2 $&$ 0  $\\
$0  $&$ 0  $&$ 0  $&$ 0  $&$ 0  $&$ 0  $&$ 0  $&$ 0 $&$ 2  $
\end{tabular} \right]. \eea
To put things back in perspective, the data collector at this point, has access to the coded symbols
\[
\underline{u}^t C_3
\]
associated with the three parity nodes. From the nature of the matrix it is evident that message symbols $u_1$, $u_5$ and $u_9$ are now available to the data-collector, and their effect can be subtracted from the remaining symbols to obtain the matrix
\beq [u_2 \ u_3 \ u_4 \ u_6 \ u_7 \ u_8]
\underbrace{\left[ \begin{tabular}{cccccc}
$ 2  $&$ 0  $&$ 1  $&$ 0  $&$ 0  $&$ 0 $\\
$ 0  $&$ 2  $&$ 0  $&$ 0  $&$ 1  $&$ 0 $\\
$ 1  $&$ 0  $&$ 2  $&$ 0  $&$ 0  $&$ 0 $\\
$ 0  $&$ 0  $&$ 0  $&$ 2  $&$ 0  $&$ 1 $\\
$ 0  $&$ 1  $&$ 0  $&$ 0  $&$ 2  $&$ 0 $\\
$ 0  $&$ 0  $&$ 0  $&$ 1  $&$ 0  $&$ 2 $\\
\end{tabular} \right]}_{C_4} \label{eq:invertNonsysEnd}.\eeq
As $2^2 \neq 1$ in $\mathbb{F}_7$, the matrix $C_4$ above can be verified to be non-singular and thus the remaining message symbols can also be recovered by inverting $C_4$.
\end{IEEEproof}

(d) \textit{One systematic node and two parity nodes:} Suppose the data-collector connects to systematic node $1$ and parity nodes $4$ and $5$. All symbols of node $1$, i.e., $\underline{u}_1$ are available to the data-collector. Thus, it needs to decode the message-vector components  $\underline{u}_2$ and $\underline{u}_3$ which are encoded using a matrix $B_1$ given by
\beq B_1 = \begin{bmatrix}
            G_2^{(4)} & G_2^{(5)} \\
 G_3^{(4)}& G_3^{(5)}
           \end{bmatrix}
\eeq

~

\textit{Claim 2:}  The block-matrix $B_1$ above is non-singular and in this way, the message-vector components  $\underline{u}_2$ and $\underline{u}_3$ can be recovered.

\begin{IEEEproof}
Once again, we begin by permuting the columns of $B_1$. For $i=2,3,1$ (in this order), we group the $i^{th}$ columns of the two parity nodes together to give the matrix
\beq B_2 =
\left[ \hspace{-.3cm}\resizebox{7.5cm}{!}{
\renewcommand{\arraystretch}{1.3}
\begin{tabular}{cc|cc|cc}
$ 2{\psi}_1^{(4)} $&$ 2{\psi}_1^{(5)} $&$0   $&$ 0          $&$ {\psi}_2^{(4)}  $&$ {\psi}_2^{(5)}$ \\
$ 2{\psi}_2^{(4)}  $&$ 2{\psi}_2^{(5)} $&$ 0 $&$ 0          $&$ 0  $&$ 0$ \\
$2{\psi}_3^{(4)}  $&$ 2{\psi}_3^{(5)}          $&$ {\psi}_2^{(4)}  $&$ {\psi}_2^{(5)} $&$ 0 $&$ 0 $ \\ \hline
$ 0          $&$ 0          $&$ 2{\psi}_1^{(4)}   $&$ 2{\psi}_1^{(5)} $&$ {\psi}_3^{(4)} $&$ {\psi}_3^{(5)}$\\
$ {\psi}_3^{(4)}  $&$ {\psi}_3^{(5)}          $&$ 2{\psi}_2^{(4)}          $&$ 2{\psi}_2^{(5)}   $&$ 0          $&$ 0 $\\
$ 0  $&$ 0          $&$ 2{\psi}_3^{(4)}          $&$ 2{\psi}_3^{(5)}   $&$ 0          $&$ 0 $
\end{tabular}} \right]. \eeq

\noindent Let $\Psi_2$ be the $(2 \times 2)$ sub-matrix of the Cauchy matrix $\Psi_3$, given by \beq{\Psi}_2 = \left[  \resizebox{!}{!}{\begin{tabular}{cc}
${\psi}_2^{(4)}$ & ${\psi}_2^{(5)}$ \\
${\psi}_3^{(4)}$ & ${\psi}_3^{(5)}$
\end{tabular}} \right]. \eeq
Since every sub-matrix of $\Psi_3$ is non-singular, so is $\Psi_2$.   Keeping in mind the fact that the data collector can perform any linear operation on the columns of $B_2$, we next multiply the last two columns of $B_2$  by ${\Psi}_2^{-1}$ (while leaving the other $4$ columns unchanged) to obtain the matrix
\beq B_3 = \left[ \resizebox{!}{!}{\begin{tabular}{cc|cc|cc}
$ 2{\psi}_1^{(4)} $&$ 2{\psi}_1^{(5)} $&$0   $&$ 0          $&$ 1 $&$ 0$\\
$ 2{\psi}_2^{(4)}  $&$ 2{\psi}_2^{(5)} $&$ 0 $&$ 0          $&$ 0  $&$ 0$ \\
$2{\psi}_3^{(4)}  $&$ 2{\psi}_3^{(5)}          $&$ {\psi}_2^{(4)}  $&$ {\psi}_2^{(5)} $&$ 0 $&$ 0 $ \\ \hline
$ 0          $&$ 0          $&$ 2{\psi}_1^{(4)}   $&$ 2{\psi}_1^{(5)} $&$0$&$1$\\
$ {\psi}_3^{(4)}  $&$ {\psi}_3^{(5)}          $&$ 2{\psi}_2^{(4)}          $&$ 2{\psi}_2^{(5)}   $&$ 0          $&$ 0 $\\
$ 0  $&$ 0          $&$ 2{\psi}_3^{(4)}          $&$ 2{\psi}_3^{(5)}   $&$ 0          $&$ 0 $
\end{tabular}} \right]~.
\eeq
The message symbols associated to the last last two columns of $B_2$ are now available to the data-collector and their effect on the rest of the encoded symbols can be subtracted out to get
\beq B_4
=\left[ \resizebox{5.3cm}{!}{\begin{tabular}{cc|cc}
$ 2{\psi}_2^{(4)}  $&$ 2{\psi}_2^{(5)} $&$ 0 $&$0$\\
$2{\psi}_3^{(4)}  $&$ 2{\psi}_3^{(5)}$&$ {\psi}_2^{(4)}$&$ {\psi}_2^{(5)} $\\ \hline
$ {\psi}_3^{(4)}  $&$ {\psi}_3^{(5)}$&$ 2{\psi}_2^{(4)}$&$ 2{\psi}_2^{(5)}   $\\
$ 0  $&$ 0          $&$ 2{\psi}_3^{(4)}$&$ 2{\psi}_3^{(5)}$
\end{tabular}} \right]~.\eeq
Along the lines of the previous case, the matrix $B_4$ above can be shown to be non-singular.  We note that this condition is equivalent to the reconstruction in a MISER code with $k=2$ and a data-collector that attempts to recover the data by connecting to the two parity nodes.
\end{IEEEproof}

\subsection{The General MISER Code for $n = 2k,~d=n-1$} \label{sec:MISER_gen}

In this section, the construction of MISER code for the general parameter set $n = 2k,~d=n-1$ is provided.  Since the MISER code is built to satisfy the cut-set bound, we have that $d=\alpha+k-1$ which implies that
\beq k=\alpha~. \eeq
This relation will play a key role in the design of generator matrices for the parity nodes as this will permit each parity node to reserve $\alpha=k$ symbols associated to linearly independent global kernels for the repair of the $k$ systematic nodes. In the example just examined, we had $\alpha=k=3$.  The construction of the MISER code for the general parameter set $n = 2k,~d=n-1$ is very much along the lines of the construction of the example code.

\subsubsection{Design of Nodal Generator Matrices}
~

The first $k$ nodes are systematic and store the message symbols in uncoded form. Thus the component generator matrices $G^{(\ell)}_i $, $1 \leq i \leq k$ of the $\ell$th systematic node, $1 \leq \ell \leq k$, are given by
\bea
G^{(\ell)}_i =
\left \lbrace \begin{array}{ll}
         I_{\alpha}  &\text{if } i=\ell \\
0_\alpha &\text{if } i\neq \ell
        \end{array} \right.
\label{eq:explicitSystematicGenMxs}.
\eea

Let $\Psi$ be an $\left(\alpha  \times (n-k)\right)$ matrix with entries drawn from $\mathbb{F}_q$ such that every sub-matrix of $\Psi$ is of full rank.   Since $n-k=\alpha=k$, we have that $\Psi$ is a square matrix \footnote{In Section~\ref{sec:connect_to_all_systematic}, we extend the construction to the even more general case of arbitrary $n$, $d \geq 2k-1$, under the added requirement however, that the replacement node connect to all of the remaining systematic nodes.  In that section, we will be dealing with a rectangular $\left(\alpha  \times (n-k)\right)$ matrix $\Psi$.}. Let the columns of $\Psi$ be given by
\bea
\Psi=\begin{bmatrix}
 \underline{\psi}^{(k+1)} & \underline{\psi}^{(k+2)} & \cdots & \underline{\psi}^{(n)}
\end{bmatrix}
\eea
where the $m$th column is given by   \beq \underline{\psi}^{(m)}=\begin{bmatrix}{\psi}^{(m)}_1 \\ \vdots \\ {\psi}^{(m)}_\alpha \end{bmatrix} . \eeq
A Cauchy matrix is an example of such a matrix, and in our construction, we will assume ${\Psi}$ to be a Cauchy matrix.

~

\begin{defn}[Cauchy matrix] An $(s \times t)$ Cauchy matrix $\Psi$ over a finite field $\mathbb{F}_q$ is a matrix whose $(i,j)$th element ($1 \leq i \leq s$, $1 \leq j \leq t$) equals $\frac{1}{(x_i-y_j)}$ where $\{x_i\}\cup\{y_j\}$ is an injective sequence, i.e., a sequence with no repeated elements.
\end{defn}

~

Thus the minimum field size required for the construction of a $(s \times t)$ Cauchy matrix is $s+t$. Hence if we choose $\Psi$ to be a Cauchy matrix,
\beq q \geq \alpha + n - k. \eeq
Any finite field satisfying this condition will suffice for our construction.
Note that since $n-k \geq \alpha \geq 2$, we have $q \geq 4$.

~

We introduce some additional notation at this point. Denote the $j$th column of the $(\alpha \times \alpha)$ matrix $G^{(m)}_i$ as $\underline{g}^{(m)}_{i,j}$, i.e., \beq G^{(m)}_i = \left[\underline{g}^{(m)}_{i,1}\quad \cdots\quad \underline{g}^{(m)}_{i,\alpha}\right].\eeq

The code is designed assuming a regeneration algorithm under which each of the $\alpha$ parity nodes passes its $\ell^{\text{th}}$ column for repair of the $\ell^{\text{th}}$ systematic node.  With this in mind, for $k+1 \leq m \leq n$, $1 \leq i,j \leq \alpha$, we choose
\beq
\underline{g}^{(m)}_{i,j}=
\left\lbrace \begin{array}{ll}
\epsilon \underline{\psi}^{(m)} &\text{if }i = j \\
{\psi}^{(m)}_i\underline{e}_j\;\; &\text{if }i\neq j
\end{array}
\right.\label{eq:choose_g}
\eeq
where $\epsilon$ is an element from $\mathbb{F}_q$ such that $\epsilon \neq 0 $ and $\epsilon^2 \neq 1$ ~(in the example provided in the previous section, $\epsilon \in \mathbb{F}_7$ was set equal to  $2$). The latter condition $\epsilon^2 \neq 1$ is needed during the reconstruction process, as was seen in the example.  Note that there always exists such a value $\epsilon$ as long as $q \geq 4$.

As in the example, the generator matrix is also designed keeping in mind the need for interference alignment. This property is utilized in the exact-repair of systematic nodes, as described in the next section.

~

\subsubsection{Exact-Repair of Systematic Nodes}

The repair process we associate with the MISER code is simple. The repair of a failed systematic node, say node $\ell$, involves each of the remaining $d=n-1$ nodes passing their $\ell$th symbols (or equivalently, associated global kernels) respectively. In the set of $\alpha$ vectors passed by the parity nodes, the $\ell$th (desired) component is independent, and the remaining (interference) components are aligned. The interference components are cancelled using the vectors passed by the remaining systematic nodes.  Independence in the desired component then allows for recovery of the desired message symbols.

The next theorem describes the repair algorithm in greater detail.

\begin{thm}
 In the MISER code, a failed systematic node can be exactly repaired by downloading one symbol from each of the remaining $d=n-1$ nodes.
\end{thm}

\begin{IEEEproof}
Consider repair of the systematic node $\ell$. Each of the remaining  $(n-1)$ nodes passes its $\ell$th column, so that the replacement node has access to the global kernels represented by the columns shown below:
\[
\left[
\renewcommand{\arraystretch}{1.43}
\begin{tabular}{>{$}c<{$}|>{$}c<{$}|>{$}c<{$}|>{$}c<{$}|>{$}c<{$}|>{$}c<{$}|>{$}c<{$}|>{$}c<{$}|>{$}c<{$}}
\underline{e}_{\ell}  &\cdots &  \underline{0} &  \underline{0} &\cdots &  \underline{0} &
{\psi}^{(k+1)}_1\underline{e}_{\ell} & \cdots & {\psi}^{(n)}_1\underline{e}_{\ell} \\

\vdots &  \ddots & \vdots & \vdots & \ddots & \vdots & \vdots & \ddots & \vdots \\

\underline{0}  &\cdots &  \underline{e}_{\ell} &  \underline{0} &\cdots &  \underline{0} &
{\psi}^{(k+1)}_{\ell-1}\underline{e}_{\ell} & \cdots &{\psi}^{(n)}_{\ell-1}\underline{e}_{\ell} \\

\underline{0} &\cdots & \underline{0}  &  \underline{0} &\cdots &  \underline{0} &
\textcolor{blue}{\epsilon \underline{\psi}^{(k+1)} }& \textcolor{blue}{\cdots} & \textcolor{blue}{\epsilon \underline{\psi}^{(n)}} \\

\underline{0} &\cdots &  \underline{0} &  \underline{e}_{\ell} &\cdots &  \underline{0} &
{\psi}^{(k+1)}_{\ell+1}\underline{e}_{\ell} &  \cdots & {\psi}^{(n)}_{\ell+1}\underline{e}_{\ell} \\

\vdots  & \ddots & \vdots & \vdots & \ddots & \vdots & \vdots  & \ddots & \vdots \\

\underline{0} &\cdots &  \underline{0} &  \underline{0}  &\cdots & \underline{e}_{\ell} &
{\psi}^{(k+1)}_{k}\underline{e}_{\ell} &  \cdots & {\psi}^{(n)}_{k}\underline{e}_{\ell}\vspace{-.35cm}
 \\

\multicolumn{6}{>{$}c<{$}}{\underbrace{\hspace{4.5cm}}}&\multicolumn{3}{>{$}c<{$}}{\underbrace{\hspace{3.45cm}}}\vspace{-.1cm}\\

\multicolumn{6}{c}{From systematic nodes}&\multicolumn{3}{c}{From parity nodes}
\vspace{-1cm}
\end{tabular}
\right],
\vspace{1cm}
\]
where $\underline{e}_{\ell}$ denotes the $\ell$th unit vector of length $\alpha$ and $\underline{0}$ denotes a zero vector of length $\alpha$.

Observe that apart from the desired $\ell$th component, every other component is aligned along the vector $\underline{e}_{\ell}$.
The goal is to show that some $\alpha$ linear combinations of the columns above will give us a matrix whose $\ell$th component equals the $(\alpha \times \alpha)$ identity matrix, and has zeros everywhere else.  But this is clear from the interference alignment structure just noted in conjunction with linear independence of the $\alpha$ vectors in the desired component:
\beq \{ \underline{\psi}^{(k+1)}, ~\cdots~, \underline{\psi}^{(n)}  \} . \eeq
\end{IEEEproof}

Next, we discuss the data reconstruction property.

~

\subsubsection{Data Reconstruction~(MDS Property)} \label{sec:recon}
For reconstruction to be satisfied, a data-collector downloading all symbols stored in any arbitrary $k$ nodes should be able to recover the $B$ message symbols.  For this, we need the $(B \times B)$ matrix formed by the columnwise concatenation of any arbitrary collection of $k$ nodal generator matrices to be non-singular. The proof of this property is along the lines of the proof in the example.  For completeness, a proof is presented in the appendix.

\begin{thm}\label{thm:MISER_gen_recon}
  A data-collector connecting to any $k$ nodes in the MISER code can recover all the $B$ message symbols.
\end{thm}

\begin{IEEEproof}
Please see the Appendix.
\end{IEEEproof}

~

\begin{note}
It is easily verified that both reconstruction and repair properties continue to hold even when we choose the generator matrices of the parity nodes $\underline{g}^{(m)}_{i,j}$, $k+1 \leq m \leq n$, $1 \leq i,j \leq \alpha$ to be given by:
\beq
\underline{g}^{(m)}_{i,j}= \left\lbrace \begin{array}{l l} \Sigma_i \underline{\psi}^{(m)} &\text{if }i = j \\
{\psi}^{(m)}_i\underline{e}_j\;\; &\text{if }i\neq j
\end{array} \right.\label{eq:choose_g2}
\eeq
where $\Sigma_i = \text{diag}\{\epsilon_{i,1}~,~\ldots~,~\epsilon_{i,\alpha}\}$ is an $(\alpha \times \alpha)$ diagonal matrix satisfying
\begin{enumerate}
 \item $\epsilon_{i,j} \neq 0$, $\quad\quad~ \forall ~i,j$
 \item $\epsilon_{i,j} \, \epsilon_{j,i} \neq 1$, $\quad \forall ~i \neq j$.
\end{enumerate}

~

\noindent
The first condition suffices to ensure exact-repair of systematic nodes.  The two conditions together ensure that the~(MDS) reconstruction property holds as well.
\end{note}

\subsection{The MISER Code for $n \geq 2k,~d=n-1$}
In this section we show how the MISER code construction for $n = 2k,~d=n-1$ can be extended to the more general case $n \geq 2k, \ d=n-1$. From the cut-set bound~\eqref{eq:MSR_beta1_parameters}, for this parameter regime, we get
\beq k \leq \alpha~. \eeq

We begin by first showing how an incremental change in parameters is possible.

~
\begin{thm} \label{thm:smaller_k}
An $[n,\ k, \ d]$, linear, systematic, exact-repair MSR code ${\cal C}$ can be derived from an $[n'=n+1,k'=k+1,d'=d+1]$ linear, systematic, exact-repair MSR code ${\cal C}'$. Furthermore if $d'=a k'+b$ in code ${\cal C}'$, $d=a k+b+(a-1)$ in code ${\cal C}$.
\end{thm}
\begin{IEEEproof}
We begin by noting that
 \bea
 n-k & = & n'-k' \\
 \alpha'& = & \alpha =d-k+1 \\
  B' = k'(d'-k'+1) & = & B + \alpha .
  \label{eq:B_difference}
  \eea
In essence, we use code shortening \cite{SloaneBook} to derive code ${\cal C}$ from code ${\cal C}'$.   Specification of code ${\cal C}$ requires that given a collection of $B=k \alpha$ message symbols, we identify the $\alpha$ code symbols stored in each of the $n$ nodes.
We assume without loss of generality, that in code ${\cal C}$, the nodes are numbered $1$ through $n$, with nodes $1$ through $k$ representing the systematic nodes.  We next create an additional node numbered $0$.

The encoding algorithm for code ${\cal C}$ is based on the encoding algorithm for code ${\cal C}'$.  Given a collection of $B$ message symbols to be encoded by code ${\cal C}$, we augment this collection by an additional $\alpha$ message symbols all of which are set equal to zero.   The first set of $B$ message symbols will be stored in systematic nodes $1$ through $k$ and the string of $\alpha$ zeros will be stored in node $0$.  Nodes $0$ through $k$ are then regarded as constituting a set of $k'=(k+1)$ systematic nodes for code ${\cal C}'$.   The remaining $(n-k)$ parity nodes are filled using the encoding process associated with code ${\cal C}'$ using the message symbols stored in the $k'$ nodes numbered $0$ through $k$. Note that both codes ${\cal C}$ and ${\cal C}'$ share the same number $(n-k)$ of parity nodes.

To prove the data reconstruction property of ${\cal C}$, it suffices to prove that all the $B$ message symbols can be recovered by connecting to an arbitrary set of $k$ nodes. Given a data-collector connecting to a particular set of $k$ nodes, we examine the corresponding scenario in code ${\cal C}'$ in which the data-collector connects to node $0$ in addition to these $k$ nodes. By the assumed MDS property of code ${\cal C}'$, all the $B$ message symbols along with the $\alpha$ message symbols stored in node $0$ can be decoded using the data stored these $(k+1)$ nodes. However, since the $\alpha$ symbols stored in node $0$ are all set equal to zero, they clearly play no part in the data-reconstruction process. It follows that the $B$ message symbols can be recovered using the data from the $k$ nodes (leaving aside node $0$), thereby establishing that code ${\cal C}$ possesses the required MDS data-reconstruction property.

A similar argument can be used to establish the repair property of code ${\cal C}$ as well.   Finally, we have
\bean
d' & = & a k'+b \\
\ \Rightarrow \ d+1 & = & a(k+1) + b \\
\ \Rightarrow \ d & = & a k + b + (a-1). \eean
\end{IEEEproof}

~

By iterating the procedure in the proof of Theorem~\ref{thm:smaller_k} above $i$ times we obtain:

~

\begin{cor}
\label{cor:MSR_higher_d}
An $[n,\ k, \ d]$ linear, systematic, exact-repair MSR code ${\cal C}$ can be constructed by shortening a $[n'=n+i,k'=k+i,d'=d+i]$ linear, systematic, exact-repair MSR code ${\cal C}'$. Furthermore if $d'=a k'+b$ in code ${\cal C}'$, $d=a k+b+i(a-1)$ in code ${\cal C}$.
\end{cor}

~

\begin{note}
It is shown in the sequel~(Section~\ref{subsec:equivalence}) that every linear, exact-repair MSR code can be made systematic. Thus, Theorem~\ref{thm:smaller_k} and Corollary~\ref{cor:MSR_higher_d} apply to any linear, exact-repair MSR code~(not just systematic).  In addition, note that the theorem and the associated corollary hold for general values of $[n, \ k, \ d]$ and are not restricted to the case of $d=n-1$. Furthermore, a little thought will show that they apply to linear codes ${\cal C}'$ that perform functional repair as well.
\end{note}

~

The next corollary follows from Corollary~\ref{cor:MSR_higher_d}, and the code-shortening method employed in the Theorem~\ref{thm:smaller_k}.

\begin{cor} \label{cor:MISER_code-shortening} The MISER code for $n \geq 2k, \ d=n-1$ can be obtained by shortening the MISER code for $n'=n+(n-2k), \ k'=k + (n-2k), \ d'=d+(n-2k)=n'-1$ .
\end{cor}

~

\begin{figure}
 \centering
\includegraphics[trim=0in 0.7in 0in 0in, clip=true,width=\textwidth]{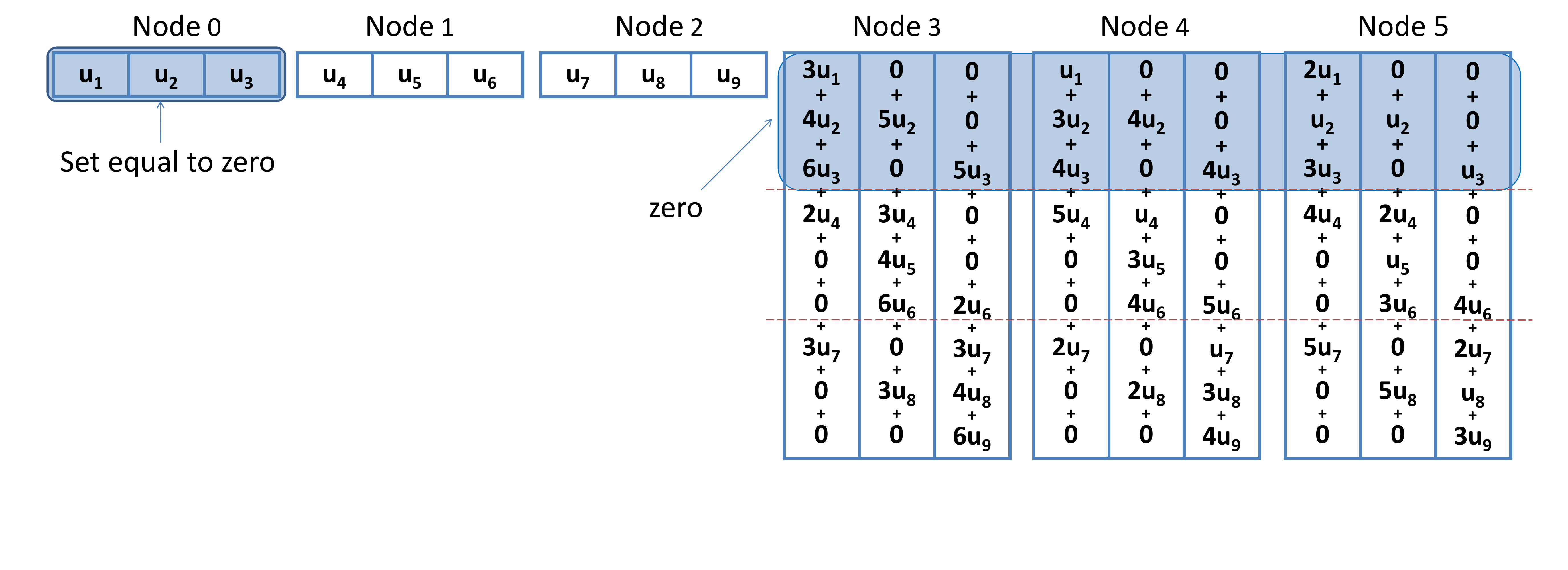}
\caption{\small
Construction of a $[n=5, \; k=2, \; d=4]$ MISER code from a $[n'=6, \; k'=3, \; d'=5]$ MISER code.  Shortening the code with respect to node zero is equivalent to removing  systematic node $0$ as well as the top component of every nodal generator matrix. The resulting $[n=5, \; k=2, \; d=4]$ MISER code has $\{u_4,\ldots,u_9\}$ as its $B=k\alpha=6$ message symbols.}
\label{fig:MISER_shorten}
\end{figure}

\textit{Example:}  The code-shortening procedure represented by Theorem~\ref{thm:smaller_k} is illustrated by the example shown in Fig.~\ref{fig:MISER_shorten}.  Here it is shown how a MISER code having code parameters $[n'=6, \; k'=3, \; d'=5]$, $\beta'=1$ and $(\alpha'=d'-k'+1=3, B'=\alpha^{'} k'=9)$ yields upon shortening with respect to the message symbols in node $0$, a MISER code having code parameters $[n=5, \; k=2, \; d=4]$, $\beta=1$ and $(\alpha=d-k+1=3, B=\alpha k=6)$.

\subsection{Extension to $ 2k-1 \leq d \leq n-1$ When The Set of Helper Nodes Includes All Remaining Systematic Nodes} \label{sec:connect_to_all_systematic}
In this section, we present a simple extension of the MISER code to the case when $2k-1 \leq d \leq n-1$, under the additional constraint however, that the set of $d$ helper nodes assisting a failed systematic node includes the remaining $k-1$ systematic nodes. The theorem below, shows that the code provided in Section~\ref{sec:MISER_gen} for $n=2k, \ d=n-1$ supports the case $d=2k-1, d \leq n-1$ as well as long as this additional requirement is met. From here on, extension to the general case $d \geq 2k-1, \ d \leq n-1$ is
straightforward via the code-shortening result in Theorem \ref{thm:smaller_k}. Note that unlike in the previous instance, the $(\alpha \times (n-k))$ Cauchy matrix used in the construction for $d < n-1$ is a rectangular matrix.

\begin{thm}
For $d=2k-1, \ d \leq n-1$, the code defined by the nodal generator matrices in equations~(\ref{eq:explicitSystematicGenMxs}) and~(\ref{eq:choose_g}),
achieves reconstruction and optimal, exact-repair of systematic nodes, provided the replacement node connects to
all the remaining systematic nodes.
\end{thm}
\begin{IEEEproof}
\textit{Reconstruction:} The reconstruction property follows directly from the reconstruction property in the case of the original code.

\textit{Exact-repair of systematic nodes:} The replacement node connects to the $(k-1)$ remaining systematic nodes and an arbitrary $\alpha$ parity nodes~(since, meeting the cut-set bound requires $d=k-1 + \alpha$). Consider a distributed storage system having only these $(k-1+\alpha)$ nodes along with the failed node as its $n$ nodes. Such a system has  $d=n-1, \ d=2k-1$ and is identical to the system described in  Section \ref{sec:MISER_gen}. Hence exact-repair of systematic nodes meeting the cut-set bound is guaranteed.
\end{IEEEproof}

\subsection{Analysis of the MISER Code}
\paragraph{Field Size Required} The constraint on the field size comes due to construction of the $\left(\alpha \times (n-k)\right)$ matrix $\Psi$ having all sub-matrices of full rank. For our constructions, since $\Psi$ is chosen to be a Cauchy matrix, any field of size $(n+d-2k+1)$ or higher suffices. For specific parameters, the matrix $\Psi$ can be handcrafted to yield smaller field sizes.

\paragraph{Complexity of Exact-Repair of Systematic Nodes} Each node participating in the exact-repair of systematic node $i$, simply passes its $i$th symbol, without any processing. The replacement node has to multiply the inverse of an~($\alpha \times \alpha$) Cauchy matrix with an $\alpha$ length vector and then perform $(k-1)$ subtractions for interference cancellation.

\paragraph{Complexity of Reconstruction} The complexity analysis is provided for the case $n=2k, \ d=n-1$, other cases follow on the similar lines. A data-collector connecting to the $k$ systematic nodes can recover all the data without any additional processing. A data-collector connecting to some $k$ arbitrary nodes has to (in the worst case) multiply the inverse of a $(k \times k)$ Cauchy matrix with $k$ vectors, along with operations having a lower order of complexity.

\subsection{Relation to Subsequent Work~\cite{Changho}}
Two regenerating codes are equivalent if one code can be transformed
into the other via a non-singular symbol remapping~(this definition
is formalized in Section~\ref{subsec:equivalence}). The capabilities
and properties of equivalent codes are thus identical in every way.

The initial presentation of the MISER code in~\cite{ourITW}~(the
name `MISER' was coined only subsequently) provided the construction
of the code along with two~(of three) parts of what may be termed as
a complete decoding algorithm, namely: (a) reconstruction by a data
collector, and (b) exact-repair of failed systematic nodes. It was
not known whether the third part of decoding, i.e., repair of a
failed parity node could be carried out by the MISER code.
Following the initial presentation of the MISER code,  the authors of \cite{Changho} show how a \textit{common eigenvector}
approach can be used to establish that exact repair of the parity nodes is also possible under the MISER code construction \footnote{In ~\cite{Changho} a class of regenerating codes is presented that have the same parameters as does the MISER code.  This class of codes can however, be shown to be equivalent to the MISER code (and hence to each other) under the equivalence notion presented in Section~\ref{subsec:equivalence}.}. 

\section{Necessity of Interference Alignment and Non-Existence of Scalar, Linear, Exact-repair MSR Codes for $d<2k-3$}\label{sec:non_exist_alpha_3}
In Section~\ref{sec:gen_explicit}, explicit, exact-repair MSR codes are constructed for the parameter regimes $(d \geq 2k-1, \ d=n-1)$ performing reconstruction and exact-repair of systematic nodes. These constructions are based on the concept of interference alignment. Furthermore, these codes have a desirable property of having the smallest possible value for the parameter $\beta$, i.e., $\beta=1$.

As previously discussed in Section~\ref{subsec:net_cod}, the problem of constructing exact-repair MSR codes is (in part) a non-multicast network coding problem. In particular, for the case of $\beta=1$, it reduces to a \textit{scalar network coding} problem. Upon increase in the value of $\beta$, the capacity of every data pipe is increased by a factor of $\beta$, thereby transforming it into a \textit{vector network coding} problem. Thus, $\beta=1$ corresponds to the absence of symbol extension, which in general, reduces the complexity of system implementation. Furthermore, as noted in Section~\ref{subsec:beta_1}, an MSR code for every larger integer value of $\beta$, can be obtained by concatenating multiple copies of a $\beta=1$ code. For this reason, the case of $\beta=1$ is of special interest and a large section of the literature in the field of regenerating codes~(\cite{WuDimISIT,ourAllerton,ourITW,DimSearch,WuArxiv,ourInterior_pts,Changho,ourProductMatrix,puyol}) is devoted to this case.

In the present section, we show that for $d<2k-3$, there exist no linear, exact-repair MSR codes achieving the cut-set bound on the repair bandwidth in the absence of symbol extension.
In fact, we show that the cut-set bound cannot be achieved even if exact-repair of only the systematic nodes is desired. We first assume the existence of such a linear, exact-repair MSR code $\mathcal{C}$ satisfying: \beq (\beta=1,\ B=k\alpha,\ \alpha=d-k+1) \eeq and \beq (d < 2k-3 \Rightarrow \alpha < k-2).\eeq
Subsequently, we derive properties that this code must necessarily satisfy. Many of these properties hold for a larger regime of parameters and are therefore of independent interest. In particular, we prove that \textit{interference alignment}, in the form described in Section~\ref{sec:intf_align}, is \textit{necessary}.  We will show that when $d <2k-3$ the system becomes over-constrained, leading to a contradiction.

We begin with some some additional notation.   

~

\begin{note}
In recent work, subsequent to the original submission of this paper, it is shown in \cite{Jafar_arxiv,Changho_arxiv_intfalign} that the MSR point under exact-repair can be achieved asymptotically for all $[n, ~k, ~d]$ via an infinite symbol extension, i.e., in the limit as $\beta \rightarrow \infty$.  This is established by presenting a scheme under which $\lim_{\beta \rightarrow \infty} \frac{\gamma}{d \beta} = 1$. Note that in the asymptotic setup, since both $\alpha, B$ are multiples of $\beta$, these two parameters tend to infinity as well.
\end{note}

\subsection{Additional Notation}\label{sec:subspaceview}
We introduce some additional notation for the vectors passed by the helper nodes to the replacement node. For $\ell,m \in \{1,\ldots,n\},\ell \neq m$, let $\underline{\gamma}^{(m,\ell)}$, denote the vector passed by node $m$ for repair of node $\ell$.  In keeping with our component notation, we will use $\underline{\gamma}^{(m,\ell)}_i$ to denote the $i$th component, $1 \leq i \leq k$, of this vector.

Recall that a set of vectors are \textit{aligned} when the vector-space spanned by them has a dimension no more than one. Given a matrix $A$, we denote its column-space by $\text{colspace}[A]$ and its~(right) null space by $\text{nullspace}[A]$. Clearly, $\underline{\gamma}^{(m,\ell)} \in \text{colspace}\left[\mathbf{G}^{(m)}\right]$.

\subsection{Equivalent Codes}\label{subsec:equivalence}
Two codes $\mathcal{C}$ and $\mathcal{C}'$ are equivalent if
$\mathcal{C}'$ can be represented in
terms of $\mathcal{C}$ by \begin{enumerate}[i)]
\item a change of basis of the vector
space generated by the message symbols~(i.e., a remapping of the message symbols), and
\item a change of basis
of the column-spaces of the nodal generator matrices~(i.e., a remapping of the symbols stored within a node). \end{enumerate}
A more rigorous definition is as follows.

~

\begin{defn}[Equivalent Codes] Two codes $\mathcal{C}$ and $\mathcal{C}'$ are equivalent if
\bea \mathbf{G}'^{(m)} &=& W \;\mathbf{G}^{(m)}\; U^{(m)} \\
\underline{\gamma}'^{(m,\ell)} &=& W \;\underline{\gamma}^{(m,\ell)} \eea
$\forall~\ell,m \in \{1,\ldots,n\},\;\ell\neq m$, for some
$(B\times B)$ non-singular matrix $W$, and some $(\alpha \times \alpha)$
non-singular matrix $U^{(m)}$.
\end{defn}

~

Since the only operator required to transform a code to its equivalent is a symbol remapping, the capabilities and properties of equivalent codes are identical in every respect. Hence, in the sequel, we will not distinguish between two equivalent codes and the notion of code equivalence will play an important role in the present section.  Here, properties of a code that is equivalent to a given code are first derived and the equivalence then guarantees that these properties hold for the given code as well. The next theorem uses the notion of equivalent codes to show that every linear exact-repair MSR code can be made systematic.

~

\begin{thm}
Every linear, exact-repair MSR code can be made systematic via a non-singular linear transformation of the rows of the generator matrix, which simply corresponds to a re-mapping of the message symbols. Furthermore, the choice of the $k$ nodes that are to be made systematic can be arbitrary.
\end{thm}
\begin{IEEEproof}

Let the generator matrix of the given linear, exact-repair MSR code $\mathcal{C}$ be $\mathbb{G}$. We will derive an equivalent code $\mathcal{C}'$ that has its first $k$ nodes in systematic form.   The reconstruction (MDS property) of code $\mathcal{C}$ implies that the $(B \times B)$ sub-matrix of $\mathbb{G}$, \[ \left[ \mathbf{G}^{(1)}~\mathbf{G}^{(2)}~\cdots~\mathbf{G}^{(k)}\right]\] is non-singular. Define an equivalent code $\mathcal{C}'$ having its generator matrix $\mathbb{G}'$ as:
\beq \mathbb{G}' = \left[ \mathbf{G}^{(1)}~\mathbf{G}^{(2)}~\cdots~\mathbf{G}^{(k)}\right]^{-1} ~ \mathbb{G}. \label{eq:convert_to_systematic}\eeq
Clearly, the $B$ left-most columns of $\mathbb{G}'$ form a $B \times B$ identity matrix, thus making the equivalent code $\mathcal{C}'$ systematic. As the repair is exact, the code will retain the systematic form following any number of failures and repairs.

The transformation in equation~\eqref{eq:convert_to_systematic} can involve any arbitrary set of $k$ nodes in $\mathcal{C}$, thus proving the second part of the theorem.
\end{IEEEproof}

~

The theorem above permits us to restrict our attention to the class of systematic codes, and assume the first $k$ nodes~(i.e., nodes $1,\ldots,k$) to be systematic. Recall that, for systematic node $\ell~(\in \{1,\ldots,k\})$,
 \bea G^{(\ell)}_i = \left \lbrace
\begin{array}{ll}
         I_{\alpha}  &\text{if } i=\ell \\
0_\alpha &\text{if } i\neq \ell
        \end{array} \right. \quad \forall i \in \{1,\ldots,k \}.
\eea
Thus, systematic node $\ell$ stores the $\alpha$ symbols in $\underline{u}_\ell$.

\subsection{Approach}
An exact-repair MSR code should be capable of performing exact-repair of any failed node by connecting to any arbitrary subset of $d$ of the remaining $(n-1)$ nodes, while meeting the cut-set bound on repair bandwidth. This requires a number of repair scenarios to be satisfied. Our proof of non-existence considers a less restrictive setting, in which exact-repair of only the systematic nodes is to be satisfied. Further, we consider only the situation where a failed systematic node is to be repaired by downloading data from a specific set of $d$ nodes, comprised of the $(k-1)$ remaining systematic nodes, and some collection of  $\alpha$ parity nodes. Thus, for the remainder of this section, we will restrict our attention to a subset of the $n$ nodes in the distributed storage network, of size $(k+\alpha)$ nodes, namely, the set of $k$ systematic nodes and the first $\alpha$ parity nodes. Without loss of generality, within this subset, we will assume that nodes $1$ through $k$ are the systematic nodes and that nodes $(k+1)$ through $(k+\alpha)$ are the $\alpha$ parity nodes.  Then with this notation, upon failure of systematic node $\ell$, $1 \leq \ell \leq k$, the replacement node is assumed to connect to nodes $\{1,\ldots,k+\alpha\}\backslash\{\ell\}$.

\begin{figure}[h]
\centering
\includegraphics[trim=0.5in 2.7in 1.5in 0.4in, clip=true, width=0.5\textwidth]{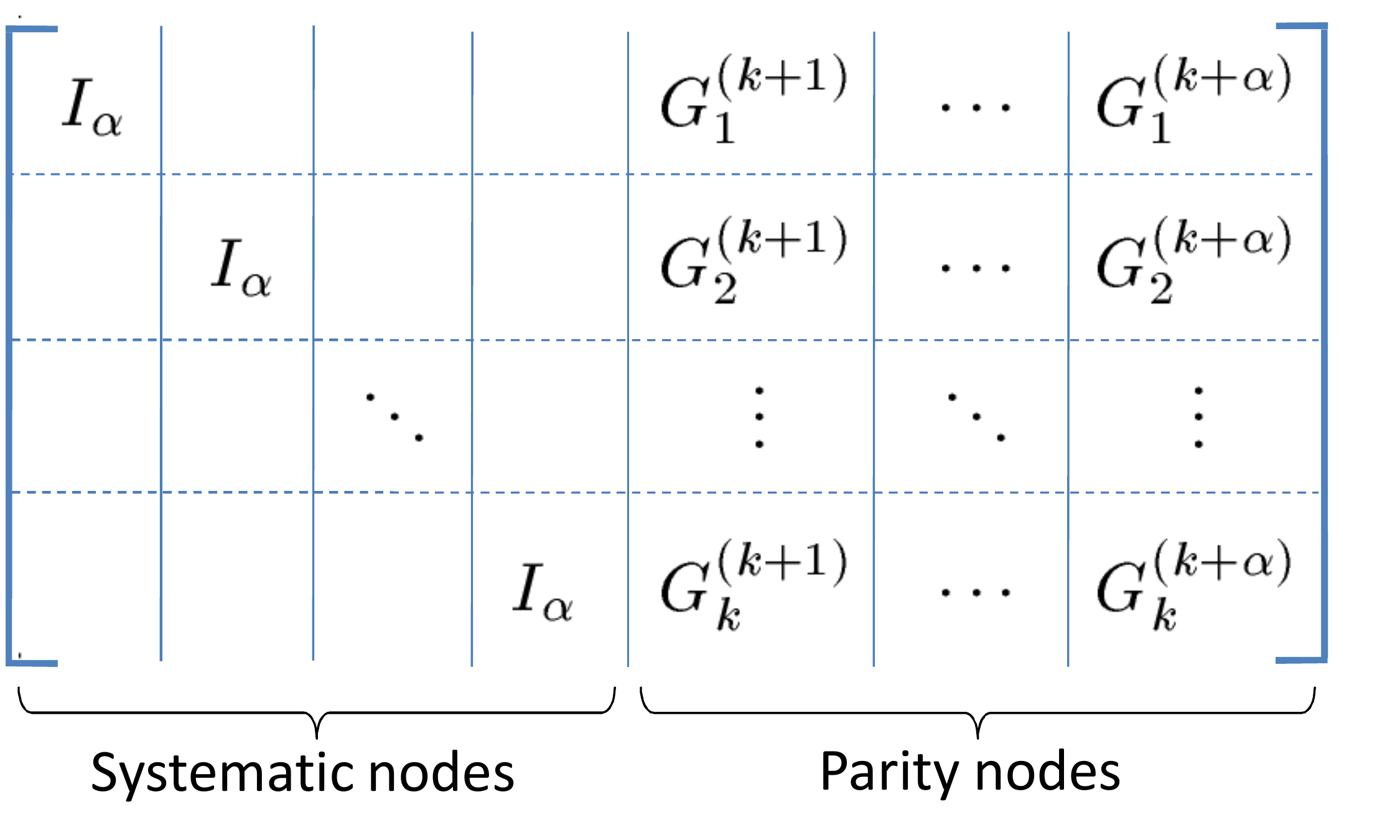}
\caption{\small The generator matrix $\mathbb{G}$ of the entire code. First $k$~(block) columns are associated with the systematic nodes $1$ to $k$ and the next $\alpha$~(block) columns to the parity nodes $(k+1)$ to $(k+\alpha)$. Empty blocks denote zero matrices.} \label{fig:non_ach_1}
\end{figure}

The generator matrix $\mathbb{G}$ of the entire code can be written in a block-matrix form as shown in Fig.~\ref{fig:non_ach_1}. In the figure, each~(block) column  represents a node and each~(block) row, a component. The first $k$ and the remaining $\alpha$ columns contain respectively, the generator matrices of the $k$ systematic nodes and the $\alpha$ parity nodes.

We now outline the steps involved in proving the non-existence
result. Along the way, we will uncover some interesting and
insightful properties possessed by linear, exact-repair MSR codes.  \ben
\item We begin by establishing that in order
to satisfy the data reconstruction property, each sub-matrix in the parity-node section
of the generator matrix~(see Fig.~\ref{fig:non_ach_1}) must be non-singular.  \item Next, we show
that the vectors passed by the $\alpha$ parity nodes for the repair
of any systematic node must necessarily satisfy two properties:
\bit \item alignment of the interference components, and \item linear independence of the desired
component. \eit
\item We then prove that in the collection of $k$ vectors passed by a
parity node for the respective repair of the $k$ systematic nodes, every $\alpha$-sized subset
must be linearly independent. This is
a key step that links the vectors stored in a node to those passed
by it, and enables us to replace the $\alpha$ columns of the
generator matrix of a parity node with the vectors it passes to aid in the repair of some
subset of $\alpha$ systematic nodes.  We will assume that these $\alpha$ systematic nodes are in fact, nodes $1$ through $\alpha$.
\item Finally, we will show that
the necessity of satisfying multiple interference-alignment
conditions simultaneously, turns out to be over-constraining, forcing alignment in the desired components as well.  This leads to a contradiction, thereby proving the non-existence result.
\een

\subsection{Deduced Properties}
\begin{pty}[Non-singularity of the Component Submatrices]\label{pty:nec_recon}
Each of the component submatrices $\{ G^{(m)}_i \mid k+1 \leq m \leq k+ \alpha, \ \ 1 \leq i \leq k \}$ is non-singular.
\end{pty}
\begin{IEEEproof}
Consider a data-collector connecting to systematic nodes $2$ to $k$
and parity node $(k+1)$.  The data-collector has thus access to the
block matrix shown in Fig.~\ref{fig:non_ach_2}.

\begin{figure}[h]
\centering
 \includegraphics[trim=0.3in 3.5in 3in 0.3in, clip=true, width=0.45\textwidth]{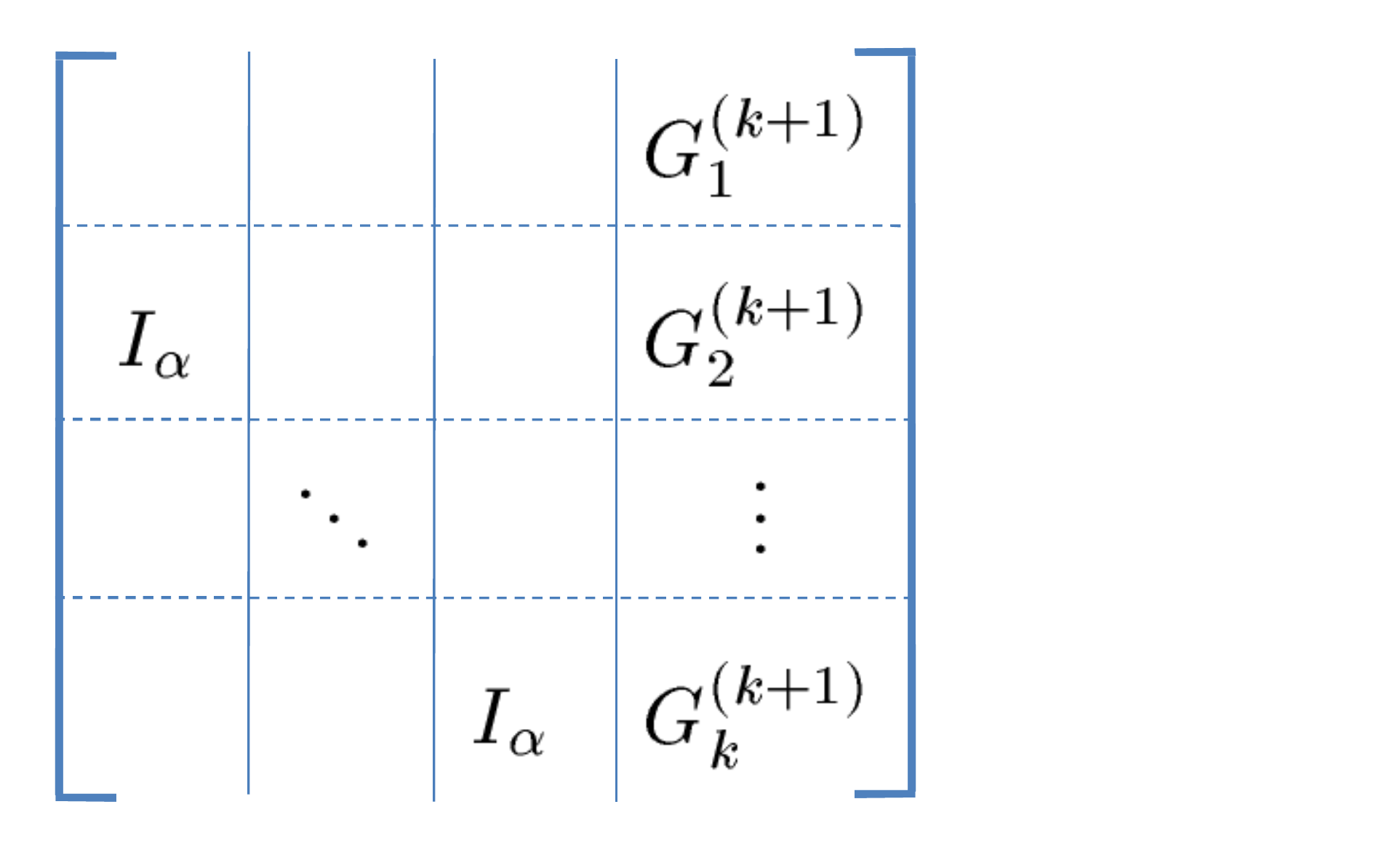}
\caption{\small The block matrix accessed by a data-collector connecting
to systematic nodes $2$ through $k$ and parity node $(k+1)$.}
\label{fig:non_ach_2}
\end{figure}

For the data-collector to recover all the data, this block matrix
must be non-singular, forcing $G_1^{(k+1)}$ to be non-singular. A similar argument shows that the same must hold in the case of each of the other component submatrices.
\end{IEEEproof}

~

\begin{cor}\label{cor:component_colspace}
Let $H=[H_1^t \ H_2^t, \cdots, H_k^t]^t$ be a $(k \alpha \times \ell)$ matrix each of whose $\ell \geq 1$ columns is a linear combination of the columns of $\mathbf{G}^{(m)}$ for some $m\in\{k+1,\ldots,k+\alpha\}$, and having $k$ components $\{H_i\}$ of size $(\alpha \times \ell)$.   Thus
\[
 \text{colspace}[H] \subseteq
\text{colspace}[\mathbf{G}^{(m)}] .
\]
Then for every $i \in
\{1,\ldots,k\}$,  we have \beq
\text{nullspace}[H_i] = \text{nullspace}[H] .  \eeq
\end{cor}

\begin{IEEEproof}
Clearly, \beq \text{nullspace}[H] \subseteq \text{nullspace}[H_i]. \eeq
Let $H = \mathbf{G}^{(m)} A$, for some $(\alpha \times \ell)$ matrix A. Then
\beq H_i = G_i^{(m)} A. \eeq
For a vector $\underline{v} \in \text{nullspace}[H_i]$,
\beq H_i \; \underline{v} = G_i^{(m)} A \; \underline{v} =\underline{0}. \eeq
However, since $G_i^{(m)}$ is of full rank~(Property~\ref{pty:nec_recon}) it follows that
\bea A \; \underline{v} &=&\underline{0} \\
\Rightarrow \ \mathbf{G}^{(m)} A \; \underline{v} &=& H \underline{v} = \underline{0} \\
\Rightarrow \ \text{nullspace}[H_i] &\subseteq& \text{nullspace}[H]. \eea
\end{IEEEproof}
The corollary says, in essence, that any linear dependence relation that holds amongst the columns of any of the components $H_i$, also extends to the columns of the entire matrix $H$ itself.

~

We next establish properties that are mandated by the repair capabilities of exact regenerating codes.  Consider the situation where a failed systematic node, say node $\ell$, \ $1 \leq \ell \leq k$, is repaired using one vector~(as $\beta=1$) from each of the remaining $k-1+\alpha$ nodes.

~

\begin{defn} When considering repair of systematic node $\ell$, $1 \leq \ell \leq k$, the $\ell$th component $\{ \underline{\gamma}^{(m,\ell)}_\ell\}$ of each of the $\alpha$ vectors $\{ \underline{\gamma}^{(m,\ell)} \mid k+1 \leq m \leq k+ \alpha \}$ passed by the $\alpha$ parity nodes  will be termed as the {\em desired component}.  The remaining
components $\{ \underline{\gamma}^{(m,\ell)}_i \mid i \neq \ell \}$ will be termed as {\em interference components}.
\end{defn}

~

The next property highlights the necessity of interference alignment in any exact-repair MSR code. Clearly, the vectors passed by the remaining $(k-1)$ systematic nodes have $\ell^{\text{th}}$ component equal to $\underline{0}$, and thus the onus of recovering the `desired' $\ell^{\text{th}}$ component of replacement node $\ell$ falls on the $\alpha$ parity nodes. However, the vectors passed by the parity nodes have non-zero `interference' components that can be nulled out only by the vectors passed by the systematic nodes. This forces an alignment in these interference components, and this is shown more formally below.

~

\begin{pty}[Necessity of Interference Alignment]\label{pty:IA_necessary} In the vectors $\{ \underline{\gamma}^{(m,\ell)} \mid k+1 \leq m \leq k+\alpha  \}$ passed by the $\alpha$ parity nodes for the repair of any systematic node (say, node $\ell$), the set of $\alpha$ interference components $\{ \underline{\gamma}^{(m,\ell)}_i \}$, $1 \leq i \leq k$, $ i \neq \ell$ must necessarily be \textit{aligned}, and the desired components $\{ \underline{\gamma}^{(m,\ell)}_\ell \}$ must necessarily be linearly independent.
\end{pty}
\begin{IEEEproof}
We assume without loss of generality that $\ell=1$, i.e., we consider repair of systematic node $1$. The matrix depicted in Fig.~\ref{fig:non_ach_3} consists of the $\alpha$ vectors needed to be recovered in the replacement node $\ell$, alongside the $d$ vectors passed by the $d$ helper nodes $2,\ldots,k+\alpha$. This matrix may be decomposed into three sub-matrices, namely:  a $(B \times \alpha)$ matrix $\Gamma_1$, comprising of the $\alpha$ columns to be recovered at the replacement node; a $(B \times (k-1))$ matrix $\Gamma_2$, comprising of the $(k-1)$ vectors passed by the remaining systematic nodes; and a $(B \times \alpha)$ matrix $\Gamma_3$,  comprising of the $\alpha$ vectors passed by the parity nodes.

\begin{figure}[h]
\centering
 \includegraphics[trim=0.4in 2.2in 0.4in 0.4in, clip=true,width=0.6\textwidth]{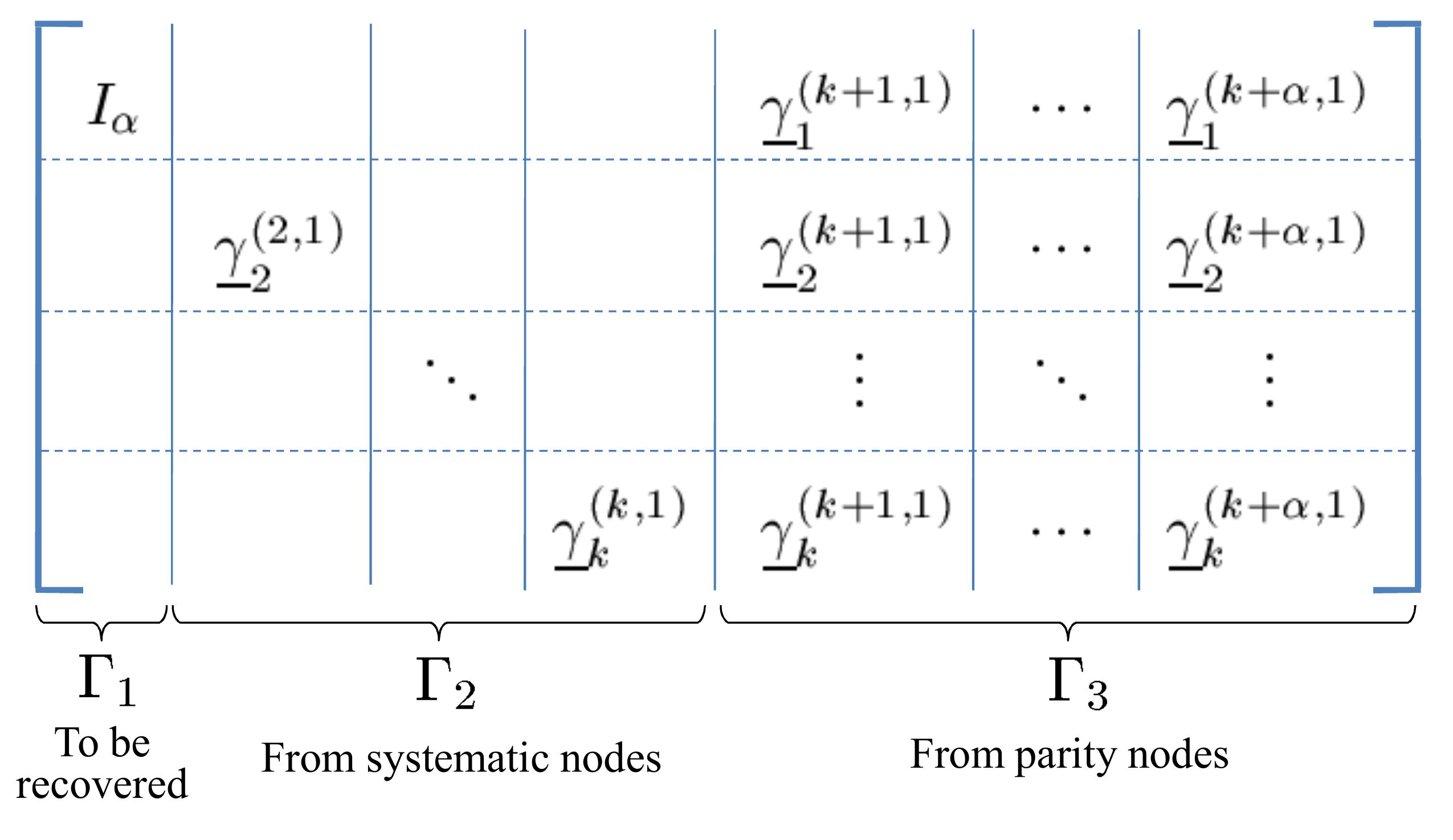}
\caption{\small Matrix depicting the $\alpha$~(global-kernel) vectors to be recovered by replacement node 1~(represented by the matrix $\Gamma_1$), alongside the $d$ vectors passed by the helper nodes $2, \ldots,k+\alpha$~(represented by $[ \Gamma_2 \mid \Gamma_3]$).
} \label{fig:non_ach_3}
\end{figure}

The vectors $\{\underline{\gamma}_1^{(k+1,1)},~ \ldots~ ,\underline{\gamma}_1^{(k+\alpha,1)}\}$ appearing in the first row of the matrix constitute the desired component; for every $i \in \{2,\ldots,k\}$, the set of vectors $\{\underline{\gamma}_i^{(k+1,1)},~ \ldots~ ,\underline{\gamma}_i^{(k+\alpha,1)}\}$, constitute interference components.  An exact-repair of node $1$ is equivalent to the recovery of $\Gamma_1$ from the columns of $\Gamma_2$ and $\Gamma_3$ through a linear transformation, and hence it must be that \beq \text{colspace} [\Gamma_1] \ \subseteq \ \text{colspace}\left[ \Gamma_2 | \Gamma_3 \right], \label{eq:IA_rk_arg_1}\eeq where `$|$' operator denotes concatenation.  When we restrict attention to the first components of the matrices, we see that we must have \beq \text{colspace}[I_{\alpha}] \ \subseteq \ \text{colspace} \left[\underline{\gamma}_1^{(k+1,1)}~ \ldots~ \underline{\gamma}_1^{(k+\alpha,1)}\right], \eeq thereby forcing the desired components $\{\underline{\gamma}_1^{(k+1,1)},~ \ldots~ ,\underline{\gamma}_1^{(k+\alpha,1)}\}$ to be linearly independent. \vspace{5pt}

Further, from \eqref{eq:IA_rk_arg_1} it follows that  \beq \text{colspace} \left[\Gamma_1 | \Gamma_2\right] \ \subseteq \ \text{colspace}\left[ \Gamma_2 | \Gamma_3 \right]. \label{eq:IA_rk_arg_2}\eeq Clearly, $\text{rank}[\Gamma_1] = \alpha$, and from Fig.~\ref{fig:non_ach_3} it can be inferred that \beq \text{rank}[\Gamma_1 | \Gamma_2]  \ = \ \alpha + \text{rank}[\Gamma_2]~.  \label{eq:IA_rk_arg_3}\eeq Moreover, as the first component in $\Gamma_3$ is of rank $\alpha$, \bea \text{rank}[\Gamma_2 | \Gamma_3] \  &\leq& \  \text{rank}[\Gamma_2] + \alpha \label{eq:IA_rk_arg_4}\\ &=&\ \text{rank}[\Gamma_1 | \Gamma_2]. \label{eq:IA_rk_arg_5}\eea It follows from equation~\eqref{eq:IA_rk_arg_2} and~\eqref{eq:IA_rk_arg_5}, that \beq \text{colspace} \left[\Gamma_1 | \Gamma_2\right] \ = \ \text{colspace}\left[ \Gamma_2 | \Gamma_3 \right], \label{eq:IA_rk_arg_6}\eeq
and this forces the interference components in $\Gamma_3$ to be aligned. Thus, for $i\in\{2,\dots,k\}$,
\beq \text{colspace}\left[\underline{\gamma}_i^{(k+1,1)}~\cdots~\underline{\gamma}_i^{(k+\alpha,1)}\right] \subseteq \text{colspace}\left[\underline{\gamma}_i^{(i,1)}\right]. \eeq

\end{IEEEproof}

~

\begin{note} Properties~\ref{pty:nec_recon} and~\ref{pty:IA_necessary} also hold for all $\beta \geq 1$, in which case, each of the $\alpha$ helper parity nodes pass a $\beta$-dimensional subspace, and each interference component needs to be confined to a $\beta$-dimensional subspace. Furthermore, the two properties also hold for all $[n,~k,~d]$ exact-repair MSR codes, when $(k-1)$ of the $d$ helper nodes along with the replacement node are viewed as systematic.
\end{note}

\begin{figure}[t]
\centering
\includegraphics[trim=0in 1.5in 0.2in 0in,clip=true,width=0.7\textwidth]{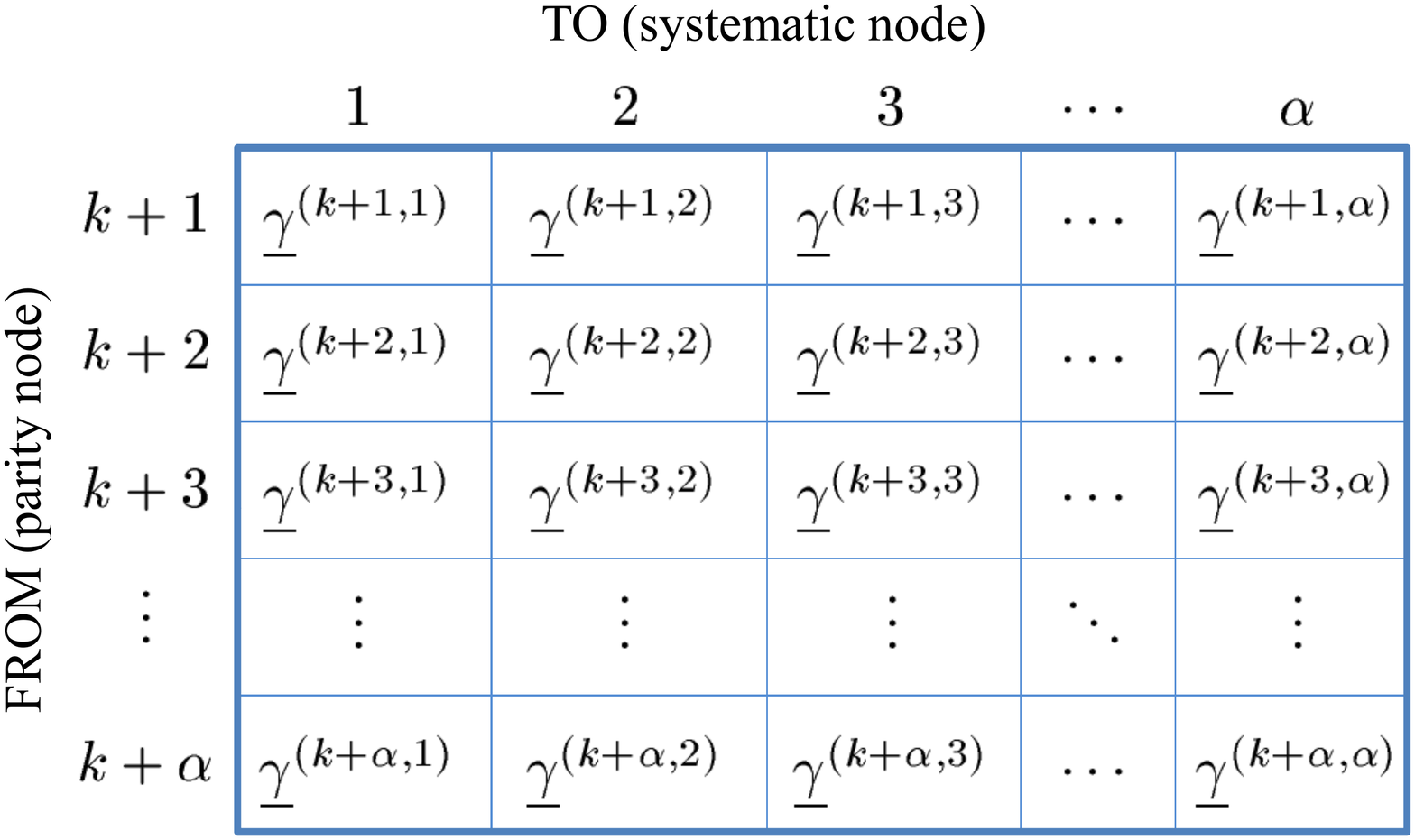}
\caption{\small Table indicating the vectors passed by the $\alpha$ parity
nodes to repair the first $\alpha$ systematic
nodes.}\label{fig:fig_fromTo}
\end{figure}

~

The next property links the vectors stored in a parity node to the vectors it passes to aid in the repair of any set of $\alpha$ systematic nodes.

~

\begin{pty}\label{pty:alpha_ind}
For $d < 2k-1$, the vectors passed by a parity node to repair any arbitrary set of $\alpha$ systematic nodes are linearly independent, i.e., for $m \in \{k+1,\ldots,k+\alpha\}$, it must be that every subset of size $\alpha$ drawn from the set of vectors \[\left\lbrace\underline{\gamma}^{(m,1)},\ldots,\underline{\gamma}^{(m,k)}\right\rbrace \] is linearly independent.  (Thus the matrix $[ \underline{\gamma}^{(m,1)}~\ldots~\underline{\gamma}^{(m,k)} ]$ may be viewed as the generator matrix of a $[k,\alpha]$-MDS code.)
\end{pty}
\begin{IEEEproof}
Consider Fig.~\ref{fig:fig_fromTo} which depicts the vectors passed
by parity nodes $\{k+1,\ldots,k+\alpha\}$  to repair systematic nodes $\{1,\ldots,\alpha\}$. From Property~\ref{pty:IA_necessary} one can infer that in column $i \in \{1,\ldots,\alpha\}$, the $i^{\text{th}}$~(desired) components of the $\alpha$ vectors are independent, and the $j^{\text{th}}$~(interference)
components for all $j \in \{1,\ldots,k\}\backslash\{i\}$ are
aligned. In particular, for all $j\in\{\alpha+1,\ldots,k\}$, the
$j^{\text{th}}$ components of each column are aligned. Note that as $d<2k-1$
we have $k>\alpha$, which guarantees that the set $\{\alpha+1,\ldots,k\}$ is non-empty and hence, the presence of an $(\alpha+1)$th component.

We will prove Property~\ref{pty:alpha_ind} by contradiction. Suppose, for example, we were to have
\beq \underline{\gamma}^{(k+1,1)} \subseteq \text{colspace}\left[\underline{\gamma}^{(k+1,2)}~\cdots~\underline{\gamma}^{(k+1,\alpha)}\right],
\eeq
which is an example situation under which the $\alpha$ vectors passed by parity node $(k+1)$ for the respective repair of the first $\alpha$ systematic nodes would fail to be linearly independent.  Restricting our attention to component $(\alpha+1)$, we get
\beq \underline{\gamma}^{(k+1,1)}_{\alpha+1} \subseteq
\text{colspace}\left[\underline{\gamma}^{(k+1,2)}_{\alpha+1}~\cdots~\underline{\gamma}^{(k+1,\alpha)}_{\alpha+1}\right]. \label{eq:non_ach_pty_3_1}\eeq
Now, alignment of component $(\alpha+1)$ along each column forces the same dependence in all other parity nodes, i.e.,
\beq \underline{\gamma}^{(m,1)}_{\alpha+1} \subseteq
\text{colspace}\left[\underline{\gamma}^{(m,2)}_{\alpha+1}~\cdots~\underline{\gamma}^{(m,\alpha)}_{\alpha+1}\right]
\quad \forall m \in \{k+2,\ldots,k+\alpha\}  \label{eq:non_ach_pty_3_2}.\eeq
Noting that a vector passed by a helper node lies in the column-space of its generator matrix, we now invoke Corollary~\ref{cor:component_colspace}:
\beq \text{nullspace}\left[\underline{\gamma}^{(m,1)}_{\alpha+1}~\cdots~\underline{\gamma}^{(m,\alpha)}_{\alpha+1}\right] = \text{nullspace}\left[\underline{\gamma}^{(m,1)}~\cdots~\underline{\gamma}^{(m,\alpha)}\right] \quad \forall m \in \{k+1,\ldots,k+\alpha\} \eeq
This, along with equations~\eqref{eq:non_ach_pty_3_1} and \eqref{eq:non_ach_pty_3_2}, implies
\beq \underline{\gamma}^{(m,1)}  \subseteq
\text{colspace}\left[\underline{\gamma}^{(m,2)}~\cdots~\underline{\gamma}^{(m,\alpha)}\right]
\quad \forall m\in \{k+1,\ldots,k+\alpha\}. \eeq
Thus the dependence in the vectors passed by one parity node carries over to every other parity node.

In particular, we have
\bea  \underline{\gamma}^{(m,1)}_1 &\subseteq&
\text{colspace}\left[\underline{\gamma}^{(m,2)}_1~\cdots~\underline{\gamma}^{(m,\alpha)}_1\right]
\quad \forall m \in \{k+1,\ldots,k+\alpha\}.
\label{eq:fromto_onecolspothers}\eea However, from
Property~\ref{pty:IA_necessary}, we know that the vectors passed to
systematic nodes $2$ to $\alpha$ have their first components
aligned, i.e., \beq \text{rank}\left[
\underline{\gamma}^{(k+1,\ell)}_1~\ldots~\underline{\gamma}^{(k+\alpha,\ell)}_1\right]
\leq 1 \qquad \forall \ell \in
\{2,\ldots,\alpha\}.\label{eq:fromto_confined}\eeq

Aggregating all instantiations~(w.r.t. $m$) of equation~\eqref{eq:fromto_onecolspothers}, the desired component is confined to:
\bea
\text{colspace}\left[\left\lbrace \underline{\gamma}^{(m,1)}_1\right\rbrace_{m=k+1}^{k+\alpha}\right] &\subseteq& \text{colspace}\left[\left\lbrace \underline{\gamma}^{(m,\ell)}_1\right\rbrace_{(m,~\ell)=(k+1,~2)}^{(k+\alpha,~\alpha)} \right]\\
\Rightarrow \text{rank}\left[\left\lbrace \underline{\gamma}^{(m,1)}_1\right\rbrace_{m=k+1}^{k+\alpha}\right] &\leq& \text{rank}\left[\left\lbrace \underline{\gamma}^{(m,\ell)}_1\right\rbrace_{(m,~\ell)=(k+1,~2)}^{(k+\alpha,~\alpha)} \right]\\
&\leq& \sum_{\ell=2}^{\alpha}\text{rank}\left[\left\lbrace \underline{\gamma}^{(m,\ell)}_1\right\rbrace_{m=k+1}^{k+\alpha} \right]\\
&\leq& \alpha-1,
\eea
where the last inequality follows from equation~\eqref{eq:fromto_confined}. This contradicts the assertion of Property~\ref{pty:IA_necessary} with respect to the desired component:
\beq \text{rank}\left[\left\lbrace \underline{\gamma}^{(m,1)}_1\right\rbrace_{m=k+1}^{k+\alpha}\right] = \ \alpha.\eeq
\end{IEEEproof}

~

\begin{note}
It turns out that an attempted proof of the analogue of this theorem for the case $\beta>1$, fails to hold.
\end{note}

~

The connection between the vectors passed by a parity node and those stored by it, resulting out of Property~\ref{pty:alpha_ind}, is presented in the following corollary.

~

\begin{cor}\label{cor:storedISpassed}
If there exists a linear, exact-repair MSR code for $d<2k-1$, then there exists an equivalent linear, exact-repair MSR code, where, for each parity node, the $\alpha$ columns of the generator matrix are respectively the vectors passed for the repair of the first $\alpha$ systematic nodes.
\end{cor}
\begin{IEEEproof}
Since a node can pass only a function of what it stores, the vectors passed by a parity node $m\in\{k+1,\ldots,k+\alpha\}$, for repair of the systematic nodes must belong to the column-space of its generator matrix, i.e.,
\beq \left[\underline{\gamma}^{(m,1)}~\cdots~\underline{\gamma}^{(m,\alpha)}\right] \subseteq \text{colspace}\left[\mathbf{G}^{(m)}\right]. \eeq
Further, Property~\ref{pty:alpha_ind} asserts that the vectors it passes for repair of the first $\alpha$ systematic nodes are linearly independent, i.e.,
\bea \text{rank}\left[\underline{\gamma}^{(m,1)}~\cdots~\underline{\gamma}^{(m,\alpha)}\right] &=& \alpha \ = \  \text{rank}\left[\mathbf{G}^{(m)}\right]. \eea

It follows that the generator matrix $\mathbf{G}^{(m)}$ is a non-singular transformation of the vectors $\left[\;\underline{\gamma}^{(m,1)}~\cdots~\underline{\gamma}^{(m,\alpha)}\;\right]$ that are passed for the repair of the first $\alpha$ systematic nodes, and the two codes with generator matrices given by the two representations are hence equivalent.
\end{IEEEproof}

~

In the equivalent code, each row of Fig.~\ref{fig:fig_fromTo} corresponds to the generator matrix $\mathbf{G}^{(m)}$ of the associated parity node, i.e.,
\beq \mathbf{G}^{(m)} = \left[\underline{\gamma}^{(m,1)} \;
\cdots  \; \underline{\gamma}^{(m,\alpha)} \right] \qquad \forall \  m\in\{k+1,\ldots,k+\alpha\}.\label{eq:nonach_storedISpassed} \eeq
Since the capabilities of a code are identical to an equivalent code, we will restrict our attention to this generator matrix for the remainder of this section. The two properties that follow highlight some additional structure in this code.

~

\begin{pty}[Code structure - what is stored] \label{pty:nonach_struct_stored}
For $d<2k-1$, any component ranging from $(\alpha+1)$ to $k$ across the generator matrices of the parity nodes differ only by the presence of a multiplicative diagonal matrix on the right, i.e.,
\beq \begin{tabular}{>{$}c<{$}>{$}c<{$}>{$}c<{$}>{$}c<{$}} G^{(k+1)}_{\alpha+1}  = H_{\alpha+1} ~\Lambda^{(k+1)}_{\alpha+1}, &G^{(k+2)}_{\alpha+1}  = H_{\alpha+1} ~\Lambda^{(k+2)}_{\alpha+1}, & \quad \cdots  \quad & G^{(k+\alpha)}_{\alpha+1}  = H_{\alpha+1}~ \Lambda^{(k+\alpha)}_{\alpha+1}\\
\vdots & \quad \vdots \quad &\quad \ddots  \quad  & \vdots \\
G^{(k+1)}_{k} \ = \ H_{k} ~ \Lambda^{(k+1)}_{k},&G^{(k+2)}_{k} \ = \ H_{k} ~ \Lambda^{(k+2)}_{k},& \quad \cdots \quad &G^{(k+\alpha)}_{k} \ = \ H_{k} ~ \Lambda^{(k+\alpha)}_{k}\end{tabular}
\label{eq:nonach_mxBelow}
 \eeq
where the matrices of the form $\Lambda_*^{(*)}$ are $\alpha \times \alpha$ diagonal matrices
(and where, for instance, we can choose $H_{\alpha+1} = G^{(k+1)}_{\alpha+1}$, in which case $\Lambda^{(k+1)}_{\alpha+1}=I_{\alpha}$).
\end{pty}
\begin{IEEEproof}
Consider the first column in Fig.~\ref{fig:fig_fromTo}, comprising of the vectors passed by the $\alpha$ parity nodes to repair node $1$. Property~\ref{pty:IA_necessary} tells us that in these $\alpha$ vectors, the components ranging from $(\alpha+1)$ to $k$ constitute interference, and are hence aligned. Clearly, the same statement holds for every column in Fig.~\ref{fig:fig_fromTo}. Thus, the respective components across these columns are aligned. Since the generator matrices of the parity nodes are as in~\eqref{eq:nonach_storedISpassed}, the result follows.
\end{IEEEproof}

~

For the repair of a systematic node, a parity node passes a vector from the column-space of its generator matrix, i.e., the vector $\underline{\gamma}^{(m,\ell)}$ passed by parity node $m$ for repair of failed systematic node $\ell$ can be written in the form:
\beq \underline{\gamma}^{(m,\ell)}~ =~ \mathbf{G}^{(m)}~ \underline{\theta}^{(m,\ell)}\eeq for some $\alpha$-length vector $\underline{\theta}^{(m,\ell)}$.

In the equivalent code obtained in~\eqref{eq:nonach_storedISpassed}, a parity node simply stores the $\alpha$ vectors it passes to repair the first $\alpha$ systematic nodes. On the other hand, the vector passed to systematic node $\ell$, $ \alpha+1 \leq \ell \leq k$, is a linear combination of these $\alpha$ vectors. The next property employs Property~\ref{pty:alpha_ind} to show that every coefficient in this linear combination is non-zero.

~

\begin{pty}[Code structure - what is passed]\label{pty:nonach_struct_passed}
For $d<2k-1$, and a helper parity node $m$ assisting a failed systematic node $\ell$\\
(a) For $\ell \in  \{1,\ldots,\alpha\}$, $\underline{\theta}^{(m,\ell)}= \underline{e}_\ell$, and\\
(b) For $\ell \in  \{\alpha+1,\ldots,k\}$, every element of $\underline{\theta}^{(m,\ell)}$ is non-zero.
\end{pty}
\begin{IEEEproof}
Part~(a) is a simple consequence of the structure of the code.   We will prove part~(b) by contradiction. Suppose $\theta^{(m,\ell)}_{\alpha}=0$, for some $\ell \in \{\alpha+1,\ldots,k\}$. Then $\underline{\gamma}^{(m,\ell)}$ is a linear combination of only the first $(\alpha-1)$ columns of $\mathbf{G}^{(m)}$. This implies,\beq \underline{\gamma}^{(m,\ell)} \subseteq \text{colspace}\left[\underline{\gamma}^{(m,1)} \cdots \underline{\gamma}^{(m,\alpha-1)} \right]. \eeq
This clearly violates Property~\ref{pty:alpha_ind}, thus leading to a contradiction.
\end{IEEEproof}

\subsection{Proof of Non-existence}
We now present the main theorem of this section, namely, the non-achievability proof. The proof, in essence, shows that the conditions of Interference Alignment necessary for exact-repair of systematic nodes, coupled with the MDS property of the code, over-constrain the system, leading to alignment in the desired components as well.

\begin{figure}[h]
\centering
\includegraphics[trim=1in 8.2in 2in 3.1in,clip=true,width=.7\textwidth]{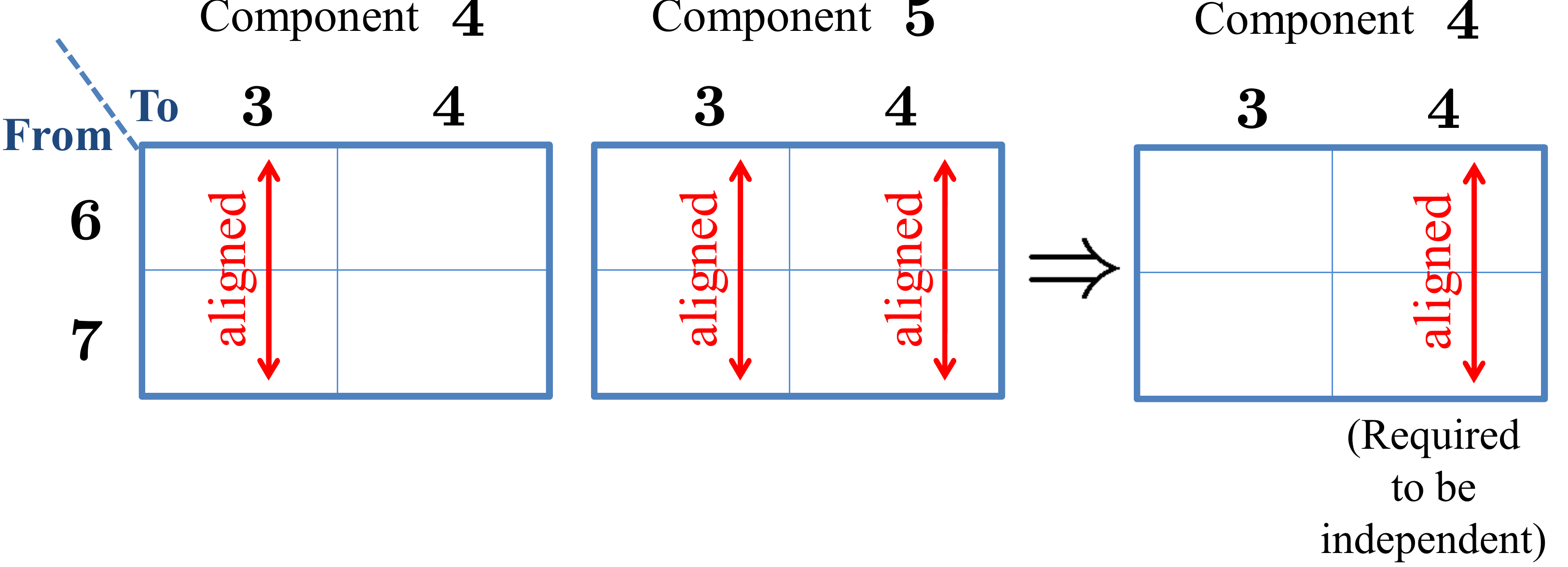}
\caption{\small A toy-example, with parameters $[n=7,~k=5,~d=6]$, to illustrate the proof of non-existence.}\label{fig:nonAch_finalProof}
\end{figure}

We begin with a toy example that will serve to illustrate the proof technique. Consider the case when $[n=7,~k=5,~d=6]$. Then it follows from \eqref{eq:MSR_beta1_parameters} that $(\alpha=d-k+1=2,~B=k \alpha=10)$. In this case, as depicted in Figure~\ref{fig:nonAch_finalProof}, in the vectors passed by parity nodes $6$ and $7$, (a) when repairing systematic node $3$, there is alignment in components $4$ and $5$, and (b) when repairing systematic node $4$, there is alignment in component $5$. It is shown that this, in turn, forces alignment in component $4$~(desired component) during repair of node $4$ which is in contradiction to the assertion of Property~\ref{pty:IA_necessary} with respect to the desired component being linearly independent.

~

\begin{thm} \label{thm:non_exist}
Linear, exact-repair MSR codes achieving the cut-set bound on the repair-bandwidth do not exist for $d<2k-3$ in the absence of symbol extension~(i.e., when $\beta=1$).
\end{thm}

\begin{IEEEproof}
Recall that achieving the cut-set bound on the repair bandwidth in the absence of symbol extension gives $d=k-1+\alpha$. For the parameter regime $d<2k-3$ under consideration, we get $k \geq \alpha+3$. Furthermore, since $\alpha >1$~\footnote{As discussed previously in Section~\ref{sec:intro}, $\alpha=1$ corresponds to a trivial scalar MDS code; hence, we omit this case from consideration.}, we have $n \geq k+2$~(as $n \geq d+1=k+\alpha$). Hence the system contains at least $(\alpha+3)$ systematic nodes and at least two parity nodes.

We use Property~\ref{pty:nonach_struct_stored} to express the generator matrix of any parity node, say node $m$, in the form:
\[ \mathbf{G}^{(m)} \ = \ \left[\begin{tabular}{>{$}c<{$}} G^{(m)}_1 \\
\vdots \\
G^{(m)}_{\alpha} \\
H_{\alpha+1} \Lambda^{(m)}_{\alpha+1} \\
\vdots \\
H_{k} \ \Lambda^{(m)}_{k}
\end{tabular}
\right].
\label{eq:nonach_nodemx}
\]

In this proof, we will use the notation $A \prec B$ to indicate that the  matrices $A$ and $B$ are scalar multiples of each other, i.e., $A$ = $\kappa B$ for some non-zero scalar $\kappa$ and write $A \nprec B$ to indicate that matrices $A$ and $B$ are \text{not} scalar multiples of each other.

We will restrict our attention to components $(\alpha+2)$ and $(\alpha+3)$. First, consider repair of systematic node $(\alpha+1)$. By the interference alignment property, Property~\ref{pty:IA_necessary},
\bea
\underline{\gamma}_{\alpha+2}^{(k+1,\alpha+1)} &\prec&  \underline{\gamma}_{\alpha+2}^{(k+2,\alpha+1)} \\
\text{i.e.,}~~~~~~~~~G^{(k+1)}_{\alpha+2} ~\underline{\theta}^{(k+1,\alpha+1)} &\prec& G^{(k+2)}_{\alpha+2}~ \underline{\theta}^{(k+2,\alpha+1)}\label{eq:nonach_final_1}\\
\Rightarrow ~ H_{\alpha+2}~ \Lambda^{(k+1)}_{\alpha+2}~ \underline{\theta}^{(k+1,\alpha+1)} &\prec& H_{\alpha+2}~ \Lambda^{(k+2)}_{\alpha+2}~ \underline{\theta}^{(k+2,\alpha+1)}\label{eq:nonach_final_2}\\
\Rightarrow ~~~~~~~~~\Lambda^{(k+1)}_{\alpha+2}~ \underline{\theta}^{(k+1,\alpha+1)} &\prec& \Lambda^{(k+2)}_{\alpha+2}~ \underline{\theta}^{(k+2,\alpha+1)}\label{eq:nonach_final_3},\eea
where, equation~\eqref{eq:nonach_final_3} uses the non-singularity of $H_{\alpha+2}$ (which is a consequence of Property~\ref{pty:nec_recon}).

We will use the notation $\Theta^{(*,*)}$ to denote  an $(\alpha \times \alpha)$ diagonal matrix, with the elements on its diagonal as the respective elements in $\underline{\theta}^{(*,*)}$. Observing that the matrices $\Lambda^{(*)}_{*}$ are diagonal matrices, we rewrite equation~\eqref{eq:nonach_final_3} as
\beq \Lambda^{(k+1)}_{\alpha+2} \Theta^{(k+1,\alpha+1)} \prec \Lambda^{(k+2)}_{\alpha+2} \Theta^{(k+2,\alpha+1)}\label{eq:nonach_final_4}.\eeq

Similarly, alignment conditions on the $(\alpha+3)$th component in the vectors passed for repair of systematic node $(\alpha+1)$ give
\beq\Lambda^{(k+2)}_{\alpha+3} \Theta^{(k+2,\alpha+1)} \prec \Lambda^{(k+1)}_{\alpha+3} \Theta^{(k+1,\alpha+1)} \label{eq:nonach_final_5},\eeq
and those on the $(\alpha+3)$th component in the vectors passed for repair of systematic node $(\alpha+2)$ give
\beq\Lambda^{(k+1)}_{\alpha+3} \Theta^{(k+1,\alpha+2)} \prec \Lambda^{(k+2)}_{\alpha+3} \Theta^{(k+2,\alpha+2)} \label{eq:nonach_final_6}.\eeq

Observe that in equations~\eqref{eq:nonach_final_4},~\eqref{eq:nonach_final_5} and \eqref{eq:nonach_final_6}, matrices $\Lambda^{(*)}_{*}$ and $\Theta^{(*,*)}$ are non-singular, diagonal matrices. As a consequence, a product~(of the terms respective in the left and right sides) of equations~\eqref{eq:nonach_final_4},~\eqref{eq:nonach_final_5} and~\eqref{eq:nonach_final_6}, followed by a cancellation of common terms leads to:
\beq \Lambda^{(k+1)}_{\alpha+2} \Theta^{(k+1,\alpha+2)} \prec \Lambda^{(k+2)}_{\alpha+2} \Theta^{(k+2,\alpha+2)}\label{eq:nonach_final_9}. \eeq
This is clearly in contradiction to Property~\ref{pty:IA_necessary}, which mandates linear independence of the desired components in vectors passed for repair of systematic node $(\alpha+2)$:
\bea H_{\alpha+2} \Lambda^{(k+1)}_{\alpha+2} \underline{\theta}^{(k+1,\alpha+2)} &\nprec& H_{\alpha+2} \Lambda^{(k+2)}_{\alpha+2} \underline{\theta}^{(k+2,\alpha+2)},\label{eq:nonach_final_7}\\
\text{i.e.},\qquad \Lambda^{(k+1)}_{\alpha+2} \Theta^{(k+1,\alpha+2)} &\nprec& \Lambda^{(k+2)}_{\alpha+2} \Theta^{(k+2,\alpha+2)}\label{eq:nonach_final_8}.
 \eea

%
%
\end{IEEEproof}

\section{Explicit Codes for $d=k+1$}\label{sec:MDSplus}
In this section, we give an explicit MSR code construction for the parameter set $\left[n,~k,~d=k+1\right]$, capable of repairing any failed node with a repair bandwidth equal to that given by the cut-set bound. This parameter set is relevant since
\begin{enumerate}[a)]
\item the total number of nodes $n$ in the system can be arbitrary~(and is not constrained to be equal to $d+1$), making the code pertinent for real-world distributed storage systems where it is natural for  the system to expand/shrink,
\item $k+1$ is the smallest value of the parameter $d$ that offers a reduction in repair bandwidth, making the code suitable for networks with low connectivity.\end{enumerate}

The code is constructed for $\beta=1$, i.e., the code does not employ any symbol extension. All subsequent discussion in this section will implicitly assume $\beta=1$.

For most values of the parameters $[n, \; k , \; d]$, $d=k+1$ falls under $d<2k-3$ regime, where we have shown (Section~\ref{sec:non_exist_alpha_3}) that exact-repair is not possible. When repair is not exact, a nodal generator matrix is liable to change after a repair process. Thus, for the code construction presented in this section, we drop the global kernel viewpoint and refer directly to the symbols stored or passed.

~

As a build up to the code construction, we first inspect the trivial case of $d=k$. In this case, the cut-set lower bound on repair bandwidth is given by
\beq d \geq k = B. \eeq
Thus the parameter regime $d=k$ mandates the repair bandwidth to be no less than the file size $B$, and has the remaining parameters satisfying \beq \left(\alpha=1, \ B=k\right). \eeq
An MSR code for these parameters is necessarily an $[n,~k]$ scalar MDS code. Thus, in this code, node $i$ stores the symbol \beq \left(\underline{p}_i^t \, \underline{u}\right), \eeq where $\underline{u}$ is a $k$-length vector containing all the message symbols, and $\lbrace\underline{r}_i\rbrace_{i=1}^{n}$ is a set of $k$-length vectors such that any arbitrary $k$ of the $n$ vectors are linearly independent. Upon failure of a node, the replacement node can connect to any arbitrary $d=k$ nodes and download one symbol each, thereby recovering the entire message from which the desired symbol can be extracted.
\begin{figure}
\centering
\includegraphics[trim=0in 0.2in 0in 0in, clip, width=\textwidth]{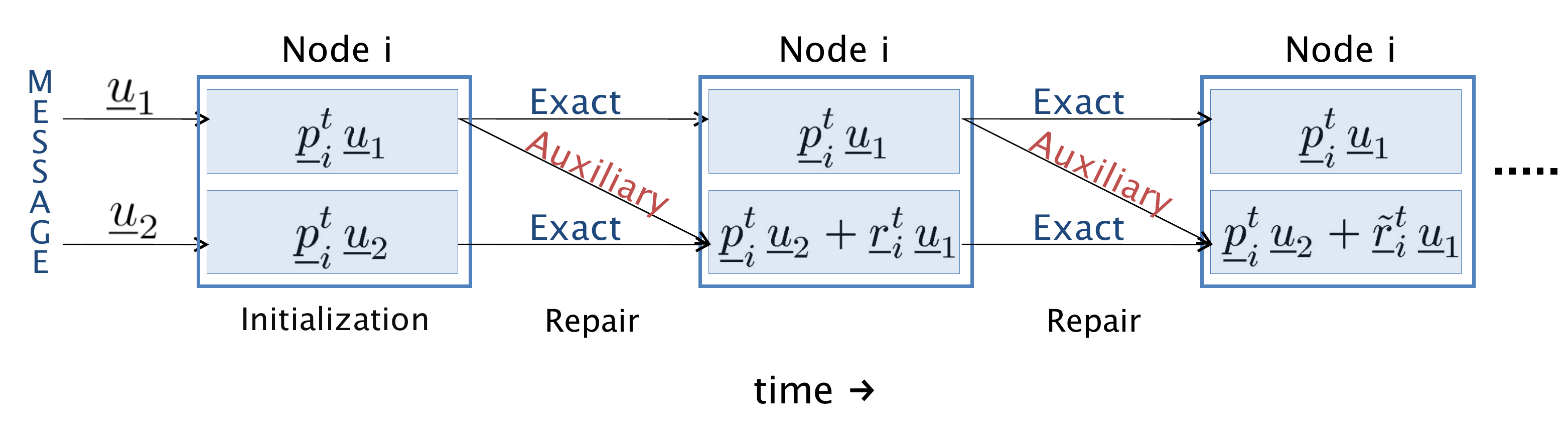}
\caption{\small Evolution of a node through multiple repairs in the MSR $d=k+1$ code.} \label{fig:dkp1_node_Evolution}
\end{figure}

When $d=k+1$, the cut-set bound~\eqref{eq:MSR_beta1_parameters} gives \beq \left( \alpha=d-k+1=2, \ B=\alpha k=2k\right) . \eeq Let the $2k$ message symbols be the elements of the $2k$-dimensional column vector \[\left[
\begin{tabular}{c}
$\underline{u}_1$\\
$\underline{u}_2$
\end{tabular}
\right], \] where $\underline{u}_1$ and $\underline{u}_2$ are $k$-length column vectors.
In the case of $d=k+1$, a code analogous to the $d=k$ code would have node $i$ storing the two symbols:
\beq \left(\underline{p}_i^t \, \underline{u}_1, ~~\underline{p}_i^t \,\underline{u}_2\right). \label{eq:dkp1_init} \eeq
Maintaining the code as in~\eqref{eq:dkp1_init}, after one or more node repairs, necessitates \textit{exact} repair of any failed node. Since in this regime, exact-repair is not possible for most values of the parameters, we allow an auxiliary component in our code, as described below.

In our construction, the symbols stored in the nodes are initialized as in~\eqref{eq:dkp1_init}. On repair of a failed node, the code allows for an auxiliary component in the second symbol. Thus, under this code, the two symbols stored in node $i,~1 \leq i \leq n$, are
\beq \text{\huge (}\underbrace{\underline{p}_i^t\,\underline{u}_1, \qquad \underline{p}_i^t \, \underline{u}_2}_{\text{Exact component}}~+\hspace{-.43cm}\underbrace{\underline{r}_i^t\,\underline{u}_1}_{\text{Auxiliary component}}\hspace{-.8cm}\text{\huge )}, \eeq where $\underline{r}_i$ is a $k$-length vector corresponding to the auxiliary component. Further, the value of $\underline{r}_i$ may alter when node $i$ undergoes repair. Hence we term this repair process as \textit{approximately-exact-repair}. For a better understanding, the system can be viewed as analogous to a $Z$-channel; this is depicted in Fig.~\ref{fig:dkp1_node_Evolution}, where the evolution of a node through successive repair operations is shown. In the latter half of this section, we will see that the set of vectors $\{\underline{r}_i\}_{i=1}^{n}$  do not, at any point in time, influence either the reconstruction or the repair process.

We now proceed to a formal description of the code construction.

\subsection{Code Construction:}
Let $\lbrace\underline{p}_i\rbrace_{i=1}^{n}$ be a set of $k$-length vectors such that any arbitrary $k$ of the $n$ vectors are linearly independent. Further, let $\{\underline{r}_i\}_{i=1}^{n}$ be a set of $k$-length vectors initialized to arbitrary values. Unlike $\lbrace\underline{p}_i\rbrace$, the vectors $\{\underline{r}_i\}$ do not play a role either in reconstruction or in repair. In our code, node $i$ stores the two symbols:
\beq \left(\underline{p}_i^t~\underline{u}_1, ~~
\underline{p}_i^t\,\underline{u}_2+\underline{r}_i^t\,\underline{u}_1\right). \eeq

Upon failure of a node, the exact component, as the name suggests, is exactly repaired. However, the auxiliary component may undergo a change. The net effect is what we term as \textit{approximately-exact-repair}.

The code is defined over the finite field $\mathbb{F}_q$ of size $q$. The sole restriction on $q$ comes from the construction of the set of vectors $\lbrace\underline{r}_i\rbrace_{i=1}^{n}$ such that every subset of $k$ vectors are linearly independent.
For instance, these vectors can be chosen from the rows of an $(n \times k)$ Vandermonde matrix or an $(n \times k)$ Cauchy matrix, in which case any finite field of size $q\geq n$ or $q \geq n+k$ respectively will suffice.

\begin{figure}[t]
\centering
\includegraphics[trim=1.1in 4.7in 2.8in 0.5in, clip, width=\textwidth]{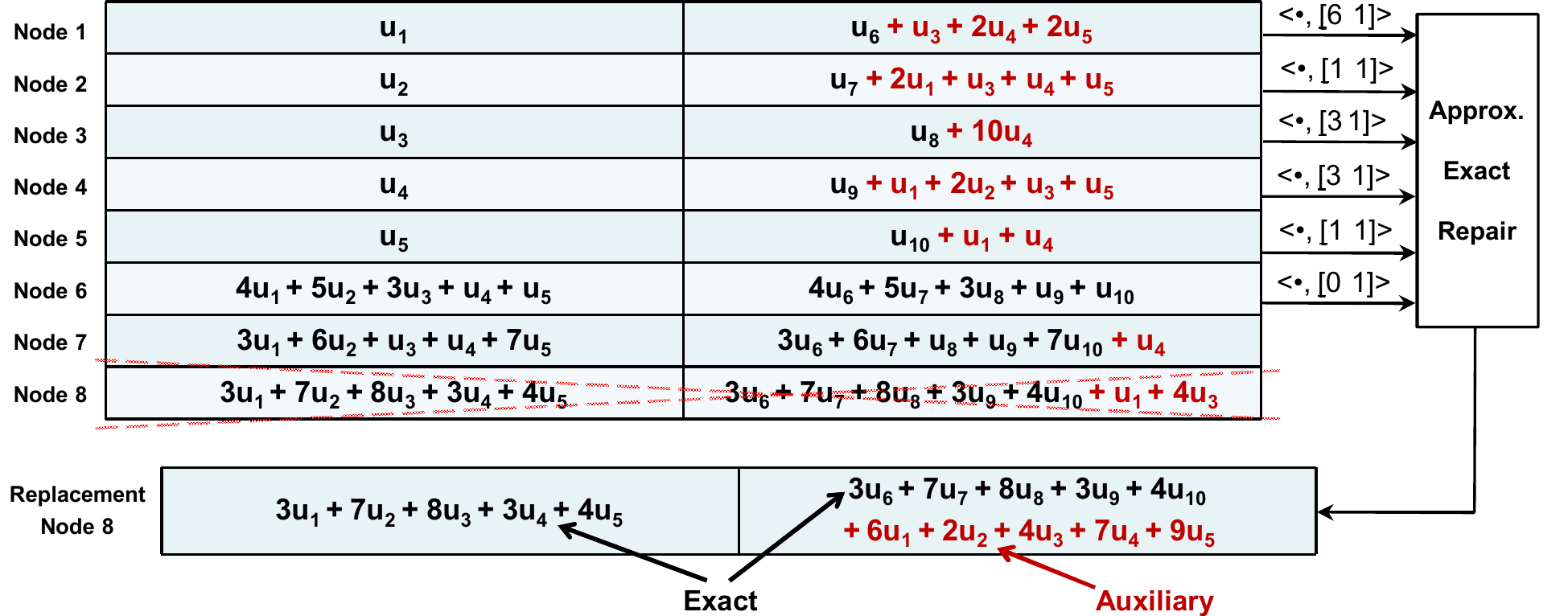}
\caption{\small A sample MSR $d=k+1$ code for the parameters $[n=8,~k=5,~d=6]$, $(\beta=1,\;\alpha=2,\;B=10)$, over $\mathbb{F}_{11}$. 
Also depicted is the repair of node $8$, assisted by helper nodes $1$ to $6$.} \label{fig:dkp1_example}
\end{figure}

\textit{Example: } Fig.~\ref{fig:dkp1_example} depicts a sample code construction over $\mathbb{F}_{11}$ for the parameters $[n=8,~k=5,~d=6]$ with $\beta=1$ giving $(\alpha=2,\ B=10)$. Here,
\[ \left[\begin{tabular}{>{$}c<{$}} \underline{p}_1^t \\
    \vdots \\
\underline{p}_8^t
   \end{tabular}\right]
 = \left[\begin{tabular}{>{$}c<{$} >{$}c<{$} >{$}c<{$} >{$}c<{$} >{$}c<{$}}
1&0&0&0&0\\
0&1&0&0&0\\
0&0&1&0&0\\
0&0&0&1&0\\
0&0&0&0&1\\
4&5&3&1&1\\
3&6&1&1&7\\
3&7&8&3&4
\end{tabular}
\right],~
 \left[\begin{tabular}{>{$}c<{$}} \underline{r}_1^t \\
    \vdots \\
\underline{r}_8^t
   \end{tabular}\right] = \left[\begin{tabular}{>{$}c<{$} >{$}c<{$} >{$}c<{$} >{$}c<{$} >{$}c<{$}}
0& 0& 1& 2& 2\\
2& 0& 1& 1& 1\\
0& 0& 0& 10&0\\
1& 2& 1& 0& 1\\
1& 0& 0& 1& 0\\
0& 0& 0& 0& 0\\
0& 0& 0& 1& 0\\
1& 0& 4& 0& 0
\end{tabular}
\right].
\]

The two theorems below show that the code described above is an $[n,~k,~d=k+1]$ MSR code by establishing respectively, the reconstruction and the repair properties of the code.

~

\begin{thm}[Reconstruction, i.e., MDS property] In the code presented, all the $B$ message symbols can be recovered by a data-collector connecting to any arbitrary $k$ nodes.\label{thm:dkp1_recon}
\end{thm}
\begin{IEEEproof}
Due to symmetry we assume (without loss of generality) that the data-collector connects to the first $k$ nodes. Then the data-collector obtains access to the $2k$ symbols stored in the first $k$ nodes:
\beq \left\lbrace\underline{p}_i^t\,\underline{u}_1, \quad
\underline{p}_i^t\,\underline{u}_2\,+\,\underline{r}_i^t\,\underline{u}_1\right\rbrace_{i=1}^{k}. \eeq

By construction, the vectors $\lbrace\underline{p}_i\rbrace_{i=1}^{k}$ are linearly independent, allowing the data-collector to recover the first message vector $\underline{u}_1$. Next, the data-collector subtracts the effect of $\underline{u}_1$ from the second term. Finally, in a manner analogous to the decoding of $\underline{u}_1$, the data-collector recovers the second message vector $\underline{u}_2$.
\end{IEEEproof}

~

\begin{thm}[Node repair] In the code presented, \textit{approximately} exact-repair of any failed node can be achieved by connecting to an arbitrary subset of $d~(=k+1)$ of the remaining $(n-1)$ nodes.\label{thm:dkp1_regen}
\end{thm}
\begin{IEEEproof}
Due to symmetry, it suffices to consider the case where helper nodes $\{1,\ldots,k+1\}$ assist in the repair of another failed node $f$. The two symbols stored in node $f$ prior to failure are \[\left(\underline{p}_{f}^t\,\underline{u}_1, \quad \underline{p}_{f}^t\,\underline{u}_2+\underline{r}_{f}^t\,\underline{u}_1\right).\] However, since repair is guaranteed to be only approximately exact, it suffices for the replacement node to obtain \[\left(\underline{p}_{f}^t\,\underline{u}_1, \quad
\underline{p}_{f}^t\,\underline{u}_2+\underline{\tilde{r}}_{f}^t\,\underline{u}_1\right),\] where $\underline{\tilde{r}}_{f}$ is an arbitrary vector that need not be identical to $\underline{r}_{f}$.

The helper nodes $\{1,\ldots,k+1\}$ pass one symbol each, formed by a linear combination of the symbols stored in them. More specifically, helper node $i, \, 1 \leq i \leq k+1$, under our repair algorithm, passes the symbol  \beq\lambda_i\left(\underline{p}_{i}^t\,\underline{u}_1\right) \,+\,
\left( \underline{p}_{i}^t\,\underline{u}_2+\underline{r}_{i}^t\,\underline{u}_1\right). \eeq

We introduce some notation at this point. For $\ell \in \{k, \, k+1\}$, let $P_{\ell}$ be a $( \ell \times k)$ matrix comprising of the vectors $\underline{p}_1, \ldots,\underline{p}_{\ell}$ as its $\ell$ rows respectively. Let $R_{\ell}$ be a second $(\ell \times k)$ matrix comprising of the vectors $\underline{r}_1, \ldots,\underline{r}_{\ell}$ as its $\ell$ rows respectively. Further, let $\Lambda_{\ell}=\text{diag}\{\lambda_1,\ldots,\lambda_{\ell}\}$ be an $(\ell \times \ell)$ diagonal matrix. In terms of these matrices, the $k+1$ symbols obtained by the replacement node can be written as the $(k+1)$-length vector
\beq (\Lambda_{k+1} P_{k+1} + R_{k+1} )~\underline{u}_1 + (P_{k+1})~\underline{u}_2~\label{eq:dkp1_regenvec}.\eeq
The precise values of the scalars $\{\lambda_i\}_{i=1}^{k+1}$ are derived below.

~

\paragraph*{Recovery of the First Symbol}
Let $\underline{\rho}$ be the linear combination of the received symbols that the replacement node takes to recover the first symbol that was stored in the failed node, i.e., we need \beq \underline{\rho}^t \left((\Lambda_{k+1} P_{k+1} + R_{k+1} )~\underline{u}_1 + (P_{k+1})~\underline{u}_2\right) \ = \  \underline{p}_{f}^t~\underline{u}_1.\eeq This requires elimination of $\underline{u}_2$, i.e., we need
\beq \underline{\rho}^{t} P_{k+1} = \underline{0}^t. \label{eq:dkp1_repair_1}\eeq
To accomplish this, we first choose
\beq \underline{\rho} = \left[\begin{tabular}{>{$}c<{$}} \underline{\rho}_1 \\ -1 \end{tabular}\right], \eeq
and in order to satisfy equation~\eqref{eq:dkp1_repair_1}, we set
\beq \underline{\rho}_1^{t} = \underline{p}_{k+1}^t P_k^{-1}. \label{eq:dkp1_repair_2}\eeq
Note that the $(k \times k)$ matrix $P_k$ is non-singular  by construction.

Now as $\underline{u}_2$ is eliminated, to obtain $\underline{p}_{f}^t~\underline{u}_1$, we need
\bea  \underline{\rho}^{t} \left( \Lambda_{k+1} P_{k+1} + R_{k+1}  \right) & = & \underline{p}_{f}^t \\
\Rightarrow \quad \underline{\rho}_1^{t} \left( \Lambda_{k} P_{k} + R_{k}  \right) & = & \underline{p}_f^t + \left( \lambda_{k+1} \; \underline{p}^{t}_{k+1} + \; \underline{r}^{t}_{k+1} \right). \label{eq:dkp1_repair_25}\eea
Choosing $\lambda_{k+1}=0$ and substituting the value of $\underline{\rho}^{t}_1$ from equation~\eqref{eq:dkp1_repair_2}, a few straightforward manipulations yield that choosing
\beq \Lambda_k = \left(\text{diag}\left[\underline{p}_{k+1}^t ~P_k^{-1}\right]\right)^{-1} \text{diag}\left[\left(\underline{p}_{f}^t ~-~ \underline{p}_{k+1}^t~ P_k^{-1}~R_k ~+~ \underline{r}_{k+1}^t\right) P_k^{-1}\right],\eeq
satisfies equation~\eqref{eq:dkp1_repair_25}, thereby enabling the replacement node to exactly recover the first symbol.
The non-singularity of the matrix $\text{diag}\left[\underline{p}_{k+1}^t ~P_k^{-1}\right]$ used here is justified as follows. Consider \beq \left[\underline{p}_{k+1}^t~ P_k^{-1}\right] P_k= \underline{p}_{k+1}^t ~. \eeq
Now, if any element of $\left[\underline{p}_{k+1}^t ~P_k^{-1}\right]$ is zero, it would imply that a linear combination of $(k-1)$ rows of $P_k$ can yield $\underline{p}_{k+1}^t$. However, this contradicts the linear independence of every subset of $k$ vectors in $\{\underline{p}_i\}_{i=1}^{n}$.

~

\paragraph*{Recovery of the Second Symbol}
Since the scalars $\{\lambda_i\}_{i=1}^{k+1}$ have already been utilized in the exact recovery of the first symbol, we are left with fewer degrees of freedom. This, in turn, gives rise to the presence of an auxiliary term in the second symbol.

Let $\underline{\delta}$ be the linear combination of the received symbols, that the replacement node takes, to obtain its second symbol $(\underline{p}_{f}^t~\underline{u}_2+\underline{\tilde{r}}_{f}^t~\underline{u}_1)$, i.e., we need
\beq \underline{\delta}^t \left((\Lambda_{k+1} P_{k+1} + R_{k+1} )~\underline{u}_1 + (P_{k+1})~\underline{u}_2\right) \ = \  \underline{p}_{f}^t~\underline{u}_2+\underline{\tilde{r}}_{f}^t~\underline{u}_1.\label{eq:dkp1_delta1}\eeq
Since the vector $\underline{\tilde{r}}_{f}$ is allowed to take any arbitrary value, the condition in~\eqref{eq:dkp1_delta1} is reduced to the requirement
\beq \underline{\delta}^{t} P_{k+1} = \underline{p}^{t}_f. \label{eq:dkp1_repair_3} \eeq
To accomplish this, we first choose
\beq \underline{\delta} = \left[\begin{tabular}{>{$}c<{$}} \underline{\delta}_1 \\ 0 \end{tabular}\right], \eeq
where, in order to satisfy equation~\eqref{eq:dkp1_repair_3}, we choose
\beq \underline{\delta}_1^{t} = \underline{p}_{f}^t P_k^{-1}~. \label{eq:dkp1_repair_4}\eeq
\end{IEEEproof}
In the example provided in Fig.~\ref{fig:dkp1_example}, node $8$ is repaired by downloading one symbol each from nodes $1$ to $6$. The linear combination coefficients used by the helper nodes are: \[ \left[ \lambda_1 ~\cdots~ \lambda_{6} \right] = \left[ 6~1~3~3~1~0\right]. \] The replacement node retains the exact part, and obtains a different auxiliary part, with $\tilde{\underline{r}}_{8} = \left[6~2~4~7~9\right].$

\section{Conclusions}\label{sec:conclusion}
This paper considers the problem of constructing MDS regenerating codes achieving the cut-set bound on repair bandwidth, and presents four major results. First, the construction of an explicit code, termed the MISER code, that is capable of performing data reconstruction as well as optimal exact-repair of the systematic nodes, is presented. The construction is based on the concept of interference alignment. Second, we show that interference alignment is, in fact, necessary to enable exact-repair in an MSR code. Thirdly, using the necessity of interference alignment as a stepping stone, several properties that every exact-repair MSR code must possess, are derived. It is then shown that these properties over-constrain the system in the absence of symbol extension for $d<2k-3$, leading to the non-existence of any linear, exact-repair  MSR code in this regime. Finally, an explicit MSR code for $d=k+1$, suited for networks with low connectivity, is presented. This is the first explicit code in the regenerating codes literature that does not impose any restriction on the total number of nodes $n$ in the system.

\appendix[Proof of Theorem \ref{thm:MISER_gen_recon}: Reconstruction in the MISER Code]
\label{app:fullrank}
\begin{IEEEproof}
The reconstruction property is equivalent to showing that the $(B \times B)$ matrix, obtained by columnwise concatenation of the  generator matrices of the $k$ nodes to which the data-collector connects, is non-singular.  We denote this $(B \times B)$ matrix by $D_1$. The proof proceeds via a series of linear, elementary row and column transformations of $D_1$, obtaining new $(B \times B)$ matrices at each intermediate step, and the non-singularity of the matrix obtained at the end of this process will establish the non-singularity of $D_1$.

Since we need to employ a substantial amount of notation here, we will make the connection between any notation that we introduce here with the notation employed in example presented in Section~\ref{sec:example}. This example provided the MISER code construction for the case $k=\alpha=3$, with the scalar selection $\epsilon=2$; we will track the case of reconstruction~(Section~\ref{sec:eg_recon}, case (d)) when the data-collector connects to the first systematic node~(node $1$), and the first two parity nodes~(nodes $4$ and $5$).

Let $\delta_1,\ldots,\delta_p$ be the $p$ parity nodes to which the data-collector connects. Let $\omega_1,\ldots,\omega_{k-p}$~($\omega_1 <\cdots < \omega_{k-p}$) be the $k-p$ systematic nodes to which the data-collector connects, and $\Omega_1,\ldots,\Omega_p$~($\Omega_1 <\cdots < \Omega_p$) be the $p$ systematic nodes to which it does \textit{not} connect. In terms of this notation, the  matrix $D_1$ is
\beq D_1 = \left[\mathbf{G}^{(\omega_1)}~ \cdots~\mathbf{G}^{(\omega_{k-p})}~ \mathbf{G}^{(\delta_1)}~ \cdots~\mathbf{G}^{(\delta_p)}\right].\eeq
Clearly, the sets $\{\omega_1,\ldots,\omega_{k-p}\}$ and $\{\Omega_1,\ldots,\Omega_p\}$ are disjoint.
In the example, the notation corresponds to $p=2$, $\delta_1=4$, $\delta_2=5$, $\omega_1 = 1$, $\Omega_1=2$ and  $\Omega_2=3$.

~

Since the data-collector can directly obtain the $(k-p)\alpha$ symbols stored in the $k-p$ systematic nodes it connects to, the corresponding components, i.e., components $\omega_1,\ldots,\omega_{k-p}$, are eliminated from $D_1$. Now, reconstruction is possible if the $(p\alpha \times p\alpha)$ matrix $D_2$ is non-singular, where $D_2$ is given by
\bea
D_2 &=& \begin{bmatrix}
   \mathbf{G'}^{(\delta_1)} & \mathbf{G'}^{(\delta_2)} \cdots &\mathbf{G'}^{(\delta_p)}
  \end{bmatrix} \nonumber \\
&=& \begin{bmatrix}
G^{(\delta_1)}_{\Omega_1} & G^{(\delta_2)}_{\Omega_1} & \cdots & G^{(\delta_p)}_{\Omega_1} \\
\vdots & \vdots & \ddots& \vdots \\
G^{(\delta_1)}_{\Omega_p} & G^{(\delta_2)}_{\Omega_p} & \cdots & G^{(\delta_p)}_{\Omega_p}
\end{bmatrix}.
\eea\noindent
The $(6 \times 6)$ matrix $B_1$ in the example corresponds to the matrix $D_2$ here.

The remaining proof uses certain matrices having specific structure. These matrices are defined in Table~\ref{table:appendix_notation}, along with their values in the case of the example.

\begin{table}[h!b!p!]
\caption{Notation: Matrices used in the Proof of Theorem~\ref{thm:MISER_gen_recon}}
\begin{tabular}{|l|l|l|l|}
\hline
Matrix & Dimension & Value & In the Example \\
\hline
$S$ & $\alpha \times p$ & $[S]_{i,j} = \psi_i^{(\delta_j)}\qquad~ \forall i,j$ &
$S=\left[\begin{array}{lll}
\psi_1^{(4)} & \psi_1^{(5)} \\
\psi_2^{(4)} & \psi_2^{(5)} \\
\psi_3^{(4)} & \psi_3^{(5)}
\end{array} \right]$\\

$\tilde{S}$ & $p \times p$ &  $[\tilde{S}]_{i,j} = \psi_{\Omega_i}^{(\delta_j)} \qquad~ \forall i,j$ &
$\tilde{S}=\left[ \begin{array}{ll}
\psi_2^{(4)} & \psi_2^{(5)} \\
\psi_3^{(4)} & \psi_3^{(5)}
\end{array} \right]=\Psi_2$\\

$T_{a,b}$ & $\alpha \times p$ & $a^{th}$ row as
$[\psi_{\Omega_b}^{(\delta_1)}~\ldots~\psi_{\Omega_b}^{(\delta_p)}]$, all other elements $0$ &
$T_{1,2} =
\left[  \begin{array}{lll}
\psi_3^{(4)} & \psi_3^{(5)} \\
0 & 0 \\
0 & 0
\end{array} \right]$\\

$\tilde{T}_{a,b}$ & $p \times p$ & $a^{th}$ row as $[\psi_{\Omega_b}^{(\delta_1)}~\ldots~\psi_{\Omega_b}^{(\delta_p)}]$, all other elements $0$  &
$\tilde{T}_{1,2} =
\left[  \begin{array}{ll}
\psi_3^{(4)} & \psi_3^{(5)} \\
0 & 0
\end{array} \right]$\\

$E_{a,b}$ & $\alpha \times p$ & Element at position $(a,b)$ as $1$, all other elements $0$ &
$
 E_{1,2}=
\left[  \begin{array}{ll}
0 & 1 \\
0 & 0 \\
0 & 0
\end{array} \right]$\\

$\tilde{E}_{a,b}$ & $p \times p$ & Element at position $(a,b)$ as $1$, all other elements $0$ &
$ \tilde{E}_{1,2} =
\left[  \begin{array}{ll}
0 & 1 \\
0 & 0
\end{array} \right]$ \\

\hline
\end{tabular}
\label{table:appendix_notation}
\end{table}
~

Note first that $\tilde{S}$, being a sub-matrix of the Cauchy matrix $\Psi$, is non-singular. Further, note the following relations between the matrices:
\beq T_{a,b} \; \tilde{S}^{-1} = E_{a,b} \label{eq:app_1} \eeq and
\beq \tilde{T}_{a,b} \; \tilde{S}^{-1} = \tilde{E}_{a,b}~. \label{eq:app_2} \eeq

We begin by permuting the columns of $D_2$. Group the $\Omega_1$th columns of $\{\mathbf{G'}^{(\delta_m)}\mid m=1,\ldots,p\}$ as the first $p$ columns of $D_3$, followed by $\Omega_2$th columns of $\{\mathbf{G'}^{(\delta_m)}\mid m=1,\ldots,p\}$ as the next $p$ columns, and so on. Thus, column number $\Omega_i$ of $\mathbf{G'}^{(\delta_m)}$ moves to the position $p \times (i-1) + m$. Next, group the $\omega_1$th columns of $\{\mathbf{G'}^{(\delta_m)}\mid m=1,\ldots,p\}$ and append this group to the already permuted columns, followed by the $\omega_2$th columns, and so on. Thus, column number $\omega_i$ of $\mathbf{G'}^{(\delta_m)}$ moves to the position $p^2 + p \times (i-1) + m$. Let $D_3$ be the $(p\alpha \times p\alpha)$ matrix obtained after these permutations. The $(6 \times 6)$ matrix $B_2$ in the example, corresponds to the matrix $D_3$ here.

Next, we note that there are $\alpha$ groups with $p$ columns each in $D_3$. The component-wise grouping of the rows in the parent matrix $D_2$ induces a natural grouping in $D_3$, with its rows grouped into $p$ groups of $\alpha$ rows each. Thus $D_3$ can be viewed as a block matrix, with each block of size $\alpha \times p$, and the dimension of $D_3$ being $p \times \alpha$ blocks. Now, in terms of the matrices defined in Table~\ref{table:appendix_notation}, the matrix $D_3$ can be written as
\beq  D_3 =
 \left[ \resizebox{!}{!}{
\begin{tabular}{c@{\hspace{5pt}}c@{\hspace{4pt}}c@{\hspace{5pt}}c|c@{\hspace{4pt}}c@{\hspace{5pt}}c}
$\epsilon S $&$ T_{\Omega_2,1} $&$ \cdots $&$ T_{\Omega_p,1} $&$ T_{\omega_1,1} $&$ \cdots $&$ T_{\omega_{k-p},1} $\\
$T_{\Omega_1,2} $&$ \epsilon S $&$  \cdots $&$ T_{\Omega_p,2} $&$ T_{\omega_1,2} $&$ \cdots $&$ T_{\omega_{k-p},2} $\\
$\vdots $&$ \vdots $&$ \ddots $&$ \vdots  $&$ \vdots $&$ \ddots $&$ \vdots $\\
$T_{\Omega_1,p} $&$ T_{\Omega_2,p} $&$ \cdots $&$ \epsilon S $&$ T_{\omega_1,p} $&$ \cdots $&$ T_{\omega_{k-p},p}$
\end{tabular}} \right]. \eeq

~

Next, as the data collector can perform any linear operation on the columns of $D_3$, we multiply the last $(k-p)$ block-columns~(i.e., blocks of $p$ columns each) in $D_3$ by $\tilde{S}^{-1}$ (while leaving the other block-columns unchanged). Using equation~\eqref{eq:app_1}, the resulting $p\alpha \times p\alpha$ matrix is
\beq D_4 = \left[ \resizebox{!}{!}{
\begin{tabular}{c@{\hspace{5pt}}c@{\hspace{4pt}}c@{\hspace{5pt}}c|c@{\hspace{4pt}}c@{\hspace{5pt}}c}
$\epsilon S $&$ T_{\Omega_2,1} $&$ \cdots $&$ T_{\Omega_p,1} $&$ E_{\omega_1,1} $&$ \cdots $&$ E_{\omega_{k-p},1} $\\
$T_{\Omega_1,2} $&$ \epsilon S $&$  \cdots $&$ T_{\Omega_p,2} $&$ E_{\omega_1,2} $&$ \cdots $&$ E_{\omega_{k-p},2} $\\
$\vdots $&$ \vdots $&$ \ddots $&$ \vdots  $&$ \vdots $&$ \ddots $&$ \vdots $\\
$T_{\Omega_1,p} $&$ T_{\Omega_2,p} $&$ \cdots $&$ \epsilon S $&$ E_{\omega_1,p} $&$ \cdots $&$ E_{\omega_{k-p},p}$\end{tabular}} \right]~.
\eeq
The $(6 \times 6)$ matrix $B_3$ in the example, corresponds to the matrix $D_4$ here.

Observe that in the block-columns ranging from $p+1$ to $\alpha$ of the matrix $D_4$, every individual column has exactly one non-zero element. The message symbols associated to these columns of $D_4$ are now available to the data-collector and their effect on the rest of the encoded symbols can be subtracted out to get the following $(p^2 \times p^2)$ matrix
\beq
D_5 = \left[ \resizebox{!}{!}{ \begin{tabular}{c@{\hspace{5pt}}c@{\hspace{4pt}}c@{\hspace{5pt}}c}
$\epsilon \tilde{S} $&$ \tilde{T}_{2,1} $&$ \cdots $&$ \tilde{T}_{p,1} $\\
$\tilde{T}_{1,2} $&$ \epsilon \tilde{S} $&$  \cdots $&$ \tilde{T}_{p,2}$\\
$\vdots $&$ \vdots $&$ \ddots $&$ \vdots  $\\
$\tilde{T}_{1,p} $&$ \tilde{T}_{2,p} $&$ \cdots $&$ \epsilon \tilde{S} $
\end{tabular}} \right]~.
\eeq
The matrix $D_5$ here, is the $(4 \times 4)$ matrix $B_4$ in the example. This is equivalent to reconstruction in the MISER code with the parameter $k$ equal to $p$ when a data-collector is attempting data recovery from the $p$ parity nodes. Hence, general decoding algorithms for data collection from the parity nodes alone can also be applied, as in the present case, where data collection is done partially from systematic nodes and partially from parity nodes. The decoding procedure for this case is provided below.

~

In the example detailed in case (c) of Section~\ref{sec:eg_recon}, where the data-collector connects to all three parity nodes, is related to this general case with $p=3$, $\tilde{S}=\Psi_3$ and $D_5=C_2$. We will track this case in the sequel.

The data-collector multiplies each of the $p$ block-columns in $D_5$ by $\tilde{S}^{-1}$. From equation~\eqref{eq:app_2}, the resultant $(p^2 \times p^2)$ matrix is
\beq D_6= \left[  \begin{array}{c c c c c}
\epsilon I_{p} & \tilde{E}_{2,1} & \tilde{E}_{3,1} & \cdots & \tilde{E}_{p,1} \\
\tilde{E}_{1,2} & \epsilon I_{p} & \tilde{E}_{3,2} & \cdots & \tilde{E}_{p,2} \\
\vdots & \vdots & \vdots & \ddots & \vdots \\
\tilde{E}_{1,p} & \tilde{E}_{2,p} & \tilde{E}_{3,p}  & \cdots & \epsilon I_{p} \\
\end{array} \right]. \eeq
The $(9 \times 9)$ matrix $C_3$ in the example, corresponds to the matrix $D_6$ here.

~

For $i=1,\ldots,p$, the $i$th column in the $i$th block-column contains exactly one non-zero element~(which is in the $i$th row of the $i$th block-row). It is evident that message symbols corresponding to these columns are now available to the data-collector, and their effect can be subtracted from the remaining symbols. This intermediate matrix corresponds to the $(6 \times 6)$ matrix $C_4$ in the example. Next we rearrange the resulting matrix by first placing the $i$th column of the $j$th block-column adjacent to the $j$th column of the $i$th block-column and repeating the same procedure for rows to get a $\left((p^2-p) \times (p^2-p)\right)$ matrix $D_7$ as
\beq D_7 = \left[  \begin{array}{ccccccc}
\epsilon & 1        & 0        & 0        & \cdots & 0        & 0       \\
1        & \epsilon & 0        & 0        & \cdots & 0        & 0       \\
0        & 0        & \epsilon & 1    & \cdots & 0        & 0       \\
0    & 0        & 1        & \epsilon & \cdots & 0        & 0       \\
\vdots & & \vdots & \vdots &\ddots  & \vdots & \vdots \\
0        & 0        & 0        & 0        & \cdots & \epsilon & 1    \\
0    & 0        & 0        & 0        & \cdots & 1        & \epsilon
\end{array} \right]. \eeq
This is a block diagonal matrix which is non-singular since $\epsilon^2 \neq 1$. Thus the remaining message symbols can be recovered by decoding them in pairs.

\end{IEEEproof}
\end{document}